\documentclass[12pt]{iopart}
\usepackage{graphicx}
\usepackage{amssymb}
\usepackage{subfigure}

\begin{document}
\title[Resonant Superfluidity in an Optical Lattice]{Resonant Superfluidity in an Optical Lattice}

\author{I Titvinidze$^{1}$, M Snoek$^{2}$ and W Hofstetter$^{1}$}
\address{$^{1}$ Institut f\"ur Theoretische Physik, Johann Wolfgang Goethe-Universit\"at, 60438 Frankfurt am Main, Germany}
\address{$^{2}$ Institute for Theoretical Physics, Valckenierstraat 65, 1018 XE Amsterdam, The Netherlands}
\ead{irakli@itp.uni-frankfurt.de}
\date{\today}

\begin{abstract}
We study a system of ultracold fermionic Potassium ($^{40}{\rm K}$) atoms  in a three-dimensional optical lattice in the vicinity of an $s$-wave Feshbach resonance. Close to resonance, the system is described by a multi-band Bose-Fermi Hubbard Hamiltonian. We derive an effective lowest-band Hamiltonian in which the effect of the higher bands is incorporated by a self-consistent mean-field approximation. The resulting model is solved by means of Generalized Dynamical Mean-Field Theory. In addition to the BEC/BCS crossover we find a phase transition to a fermionic Mott insulator at half filling, induced by the repulsive fermionic background scattering length. We also calculate the critical temperature of the BEC/BCS-state and find it to be minimal at resonance. 
\end{abstract}

\submitto{\NJP}

\maketitle

\section{Introduction}

The first experimental realizations  of Bose-Einstein condensates in dilute atomic gases of rubidium \cite{Anderson95}, lithium \cite{Bradley95} and sodium \cite{Davis95} atoms initiated a new field of condensed-matter research, providing an ideal laboratory for comparing theoretical models and experimental results with high accuracy. In particular the important consequences of Bose-Einstein condensation could be investigated, which up to 1995 had remained an elusive and inaccessible phenomenon in experiments. 

%In the last 15 years there has been significant experimental progress. These include accessing of hydrodynamic nature of collective oscillations \cite{Jin96,Mewes96}, the observation of the interference of matter waves \cite{Andrews97}, realization of: spinor condensates \cite{Stenger98}, Josephson like effects \cite{Cataliotti01,Albiez05}, superfluid-Mott insulator transition \cite{Greiner02, Greiner03, Trotzky08, Foelling07, DeMarco05}, Hanbury-Brown-Twiss effect \cite{Schellekens05}, and many other phenomenon. 

Not long after the first realization of BEC, ultracold fermionic gases were studied experimentally as well. The first important results of quantum degeneracy in trapped Fermi gases were obtained in 1999 at JILA \cite{Demarco99} and later on by other groups \cite{Schreck01,Truscott01}. A break-through experiment in this field was the investigation of fermionic superfluidity at the crossover between the BEC state and the Bardeen-Cooper-Schrieffer (BCS) state \cite{Greiner03, Jochim03, Zwierlein03, Bourdel04, Zwierlein04}. This is  made possible by the use of Feshbach resonances, which have become an indispensable experimental tool for ultracold atom experiments. Feshbach resonances not only allow one to tune the interatomic interaction with high precision, but also make it possible to increase it to a level where the critical temperature becomes high enough for the investigation of attraction-induced superfluidity. On the contrary, away from resonance the critical temperature is usually exponentially suppressed and experimentally inaccessible. The regime of strong, even diverging interactions, the so-called unitarity region, on the other hand defines a new field of research where standard mean-field methods break down and the physics has to be described in a non-perturbative way. Moreover, this system allows for the experimental investigation of the BEC-BCS crossover: for negative scattering lengths the system is a BCS superfluid, whereas for positive scattering length fermionic atoms with opposite spin pair up to form a bosonic molecular bound state.
More recent experimental work has focused on studying the effect of spin imbalance on the BCS state, i.e. the case when an unequal number of fermions occupies the two different spin states \cite{Zwierlein06a,Zwierlein06b,Partridge06,Shin06}, as well as mixtures of fermions with unequal masses, such as $^6{\rm Li}$ and $^{40}{\rm K}$ \cite{Taglieber08, Wille08}. 

Also the effect of periodic potentials on trapped Fermi gases  has been studied experimentally \cite{Modugno03, Stoeferle06,Chin06,Rom06}. Recently, evidence for a fermionic Mott insulator was obtained in a system of repulsively interacting $^{40}{\rm K}$ fermions in an optical lattice \cite{Joerdens08, Schneider08}. Optical lattices and Feshbach resonances have been combined experimentally as well: the group at ETH Z\"urich reported the production of $^{40}{\rm K}$ molecules in three-dimensional cubic optical lattices using $s$-wave Feshbach resonances in early 2006 \cite{Stoeferle06}, but no evidence of a superfluid state in the lattice was found until later that year, when superfluid $^6{\rm Li}$ was loaded in an optical lattice at MIT where both a condensate and an insulating state were observed \cite{Chin06}. The results where interpreted in terms of a superfluid-to-Mott insulator transition, for which a detailed theoretical description is still lacking.

\begin{figure}
\begin{center}
\includegraphics[scale=0.4]{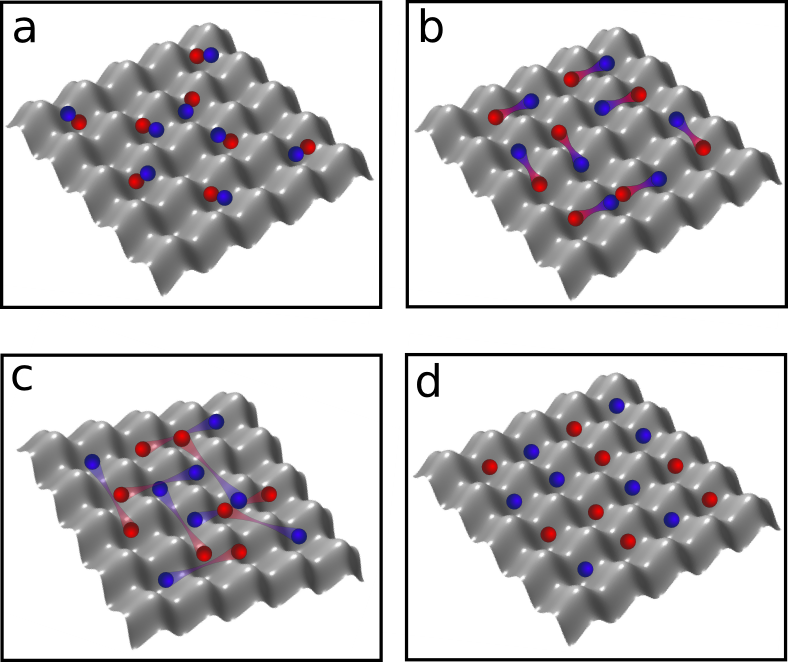}  
\end{center} 
\vspace{-0.2cm}
\caption{Schematic picture of the BEC-BCS crossover in an optical lattice. By tuning the interaction strength between the two fermionic spin states, one obtains a smooth crossover from a BEC regime of tightly bound bosonic molecules (a) to a BCS regime of large Cooper pairs (c). In between these two extremes, one encounters an intermediate crossover regime where the pair size is comparable to the interparticle spacing (b). For total fermionic filling one, the system can undergo a phase transition to the Mott insulator phase (d).}
\label{BEC-BCS}
\end{figure}

In this work we study an ultracold mixture of fermionic atoms in two different hyperfine states in a three-dimensional optical lattice close to a Feshbach resonance. This system has all the characteristics of the continuum BEC/BCS crossover described above: 
For magnetic field values below the resonance, fermions with different spin form bosonic molecules (see Fig.~\ref{BEC-BCS}).  By varying the magnetic field the bosonic level is detuned relative to the fermionic one, which changes the ratio of the densities of fermions and molecular bosons as well as the the effective interaction between the fermions. On top of this BEC-BCS crossover physics, which is familiar from the system without lattice, new features emerge when an optical lattice is applied. The most prominent one is the occurrence of a fermionic Mott insulator for half-filled fermions deep in the BCS regime, which is stabilized by the repulsive fermionic background scattering length. As described below, we find a first order transition between the BEC/BCS state and the Mott insulator. For a total filling of two fermions per site the BCS state competes against a band insulating state \cite{Zhou06, Watanabe09,Zhai07}.

%%% Dickerscheid05, I remove this citation because 
%%% In this paper they studying
%%% 1) single-band model;
%%% 2) mixture of spinless fermions and bosons which can crate fermionic molecule (bound state of fermion and boson)
%%% You can add a new sentence where you can mention that such a system is studied as well.
%%% I only commenting out it from the citation list. 

The presence of an optical  lattice allows to utilize powerful non-perturbative methods that are available for lattice systems. Here we apply Generalized Dynamical Mean-Field Theory (GDMFT) \cite{Titvinidze08, Titvinidze09} to the system described above. However, GDMFT is a single-band approach, whereas Feshbach-resonant interactions in an optical lattice lead to a multi-band model \cite{Duan06, Koetsier, Dickerscheid05b, Dickerscheid05c, Diener06, Gubbels06}. We therefore first perform a mean-field decoupling of the higher bands, thereby deriving an effective single-band Hamiltonian which is self-consistently coupled to the higher bands.

%%%%%%%%%%%%%%%%%%%%%%%%%%%%%%%%%%%%%%%%%%%%%%%%%%%%%%%%%%
%%%%%%%%%%%%%%%%%%%%%%%%%%%%%%%%%%%%%%%%%%%%%%%%%%%%%%%%%%

The paper is organized as follows: in the next section we introduce the microscopic model and in Sec. \ref{Method} we introduce the Generalized Dynamical Mean-Field approach we use to solve this model. In Sec. \ref{Results} we present the result of our numerical calculations and in Sec. \ref{Summary} we end up with concluding remarks. In the appendix, we describe in detail how the self-energy for the resonantly interacting Bose-Fermi mixture studied here can be calculated in the Dynamical Mean-Field framework.

\section{Microscopic Model}\label{Model}

Studying ultracold fermions close to a Feshbach resonance is a challenging problem. Due to the fact that exactly on resonance the scattering length is infinite, the standard fermionic Hubbard Hamiltonian cannot be defined. To solve this problem, it is necessary to formulate a two-channel Hamiltonian by separating out the resonance state and treating it explicitly \cite{Holland01}. The nonresonant contributions give rise to a background scattering length.
% and characterize the interactions between fermions. 
As the Feshbach resonance occurs due to a coupling with the bosonic molecular state, the additional degrees of freedom introduced in the two-channel theory are bosons \cite{Holland01}.  

%%% Dickerscheid05, I remove this citation the reason see comment above.

An ultracold atomic gas of fermionic atoms and molecular bosons close to a Feshbach resonance in the presence of an optical lattice is thus well described by a Bose-Fermi Hubbard model \cite{Koetsier,Carr-Holland}.  In our calculation we assume the molecular bosons to be in the lowest band. For the fermions, on the other hand, we have to take into account also the higher bands, in order to properly incorporate the two-body physics associated with the Feshbach resonance \cite{Koetsier, Dickerscheid05b, Busch}. Since the bandwidth is much smaller than the band gap, we approximate the higher bands to be flat and only take into account the full band-structure for the lowest band. Moreover, we neglect the interaction between fermions in higher bands with each other and with the bosons. This is justified because the filling in the higher bands is very small, so that interaction effects are also small. The Hamiltonian thus has the following form:
\begin{eqnarray}
\label{Hamiltonian_full}
\hat {\cal H}&=&\hat {\cal H}_f^0+\hat {\cal H}_b+\hat {\cal H}_{fb}^0+\sum_{l=1}^\infty (\hat {\cal H}_f^l+ 
\hat {\cal H}_{fb}^l) \, , \\
\label{Hamiltonian_f0}
\hat {\cal H}_f^0 &=& -t_f \sum_{\langle ij\rangle}\hat  c_{i\sigma,0}^{\dagger} \hat c_{j\sigma,0}^{\phantom\dagger} +U_f\sum_{i}\hat n_{i,\uparrow,0}^f\hat n_{i,\downarrow,0}^f %\nonumber\\
-(\mu-\frac{3\hbar\omega}{2})\sum_{i}\hat n_{i,0}^f \, , \\
\label{Hamiltonian_b}
\hat {\cal H}_b &=&-t_{b}\sum_{\langle ij\rangle}\hat b_i^\dagger \hat  b_j^{\phantom\dagger}
+\frac{U_b}{2}\sum_i \hat n_i^b (\hat n_i^b-1) %\nonumber\\
-(2\mu-\delta-\frac{3\hbar\omega}{2})\sum_i\hat n_i^b \, , \\
\label{Hamiltonian_fb0}
\hat {\cal H}_{fb}^0 &=& U_{fb} \sum_i \hat n_i^b \hat n_{i,0}^f + 
g_0 \sum_i \left(\hat b_i^\dagger \hat c_{i\uparrow,0}^{\phantom\dagger}\hat c_{i\downarrow,0}^{\phantom\dagger}+h.c\right) \, , \\
\label{Hamiltonian_fl}
\hat {\cal H}_{f}^l &=& \sum_i \left(\left(2l+\frac{3}{2} \right) \hbar\omega -\mu \right)\hat n_{i,l}^f \, , \\
\label{Hamiltonian_fl2}
\hat {\cal H}_{fb}^l &=&
g_l \sum_i \left(\hat b_i^\dagger \hat c_{i\uparrow,l}^{\phantom\dagger}\hat c_{i\downarrow,l}^{\phantom\dagger}+h.c\right) \, ,
\end{eqnarray}
where $\hat c_{i\sigma,l}^\dagger$ is the creation operator of a fermion with spin $\sigma$ for the $l$-th band on lattice site $i$. $\hat b_i^\dagger$ is the creation operator of a boson at site $i$.  $\hat n_{i\sigma,l}^f=\hat c_{i\sigma,l}^\dagger \hat c_{i\sigma,l}^{\phantom\dagger}$, $\hat n_{i,l}^f=\hat n_{i\uparrow,l}+\hat n_{i\downarrow,l}$ are the fermionic number operators, and $\hat n_i^b=\hat b_i^\dagger \hat b_i^{\phantom\dagger}$ is the bosonic number operator. $t_f$ and $t_b$ are the tunneling amplitude for fermions and bosons, respectively.  $U_{f}$, $U_b$ and $U_{fb}$ are Fermi-, Bose-, and Bose-Fermi-Hubbard interactions, respectively. These interactions arise due to the background scattering lengths. Furthermore $\mu$ is the chemical potential, $\delta$ is the detuning of the bosonic level, and $\omega$ is the frequency of the harmonic oscillator associated with an optical lattice well. Finally, $g_l=g_0\sqrt{L^{(1/2)}_{l}(0)}$ is the Feshbach coupling to the $l$-th band of the lattice,
where $g_0$ is the Feshbach coupling for the lowest Hubbard band and $L^{(1/2)}_{l}(0)$ is the generalized Laguerre  polynomial. Choosing the Feshbach couplings in this way guarantees that the two-body physics associated with the Feshbach resonance is incorporated exactly \cite{Koetsier,Busch}.

The parameters of the generalized Hubbard Hamiltonian (\ref{Hamiltonian_full}) are given by:
%%%%%%%%%%%%%%%%%%%%%%%%%%%%%%%%%%%%%%%%%%%%%%%%%%%%%%%%%%%%%%%%%%%%%%%%%%%%%%%%%%%%%%
\begin{eqnarray}
\label{J_exp}
&& t_{b(f)}\simeq  \frac{4}{\sqrt{\pi}} E_r^{b(f)} \left(\frac{V_0}{E_r^{b(f)}}\right)^{3/4}\hspace{-0.25cm}\exp\left[-2\sqrt{\frac{V_0}{E_r^{b(f)}}}\right] \, ,\\ 
\label{U_b&f_exp} 
&& U_{b(f)} \simeq  \sqrt{\frac{8}{\pi}} k a_{b(f)} E_r^{b(f)} \left(\frac{V_0}{E_r^{b(f)}}\right)^{3/4} \, , \\
\label{U_fb_exp}
&& U_{fb} \simeq  \frac{4}{\sqrt{\pi}} k a_{fb} E_r^{b} \frac{1+m_b/m_f}{(1+\sqrt{m_b/m_f})^{3/2}}\left(\frac{V_0}{E_r^{b}}\right)^{3/4} \, ,\\
%derived
\label{g_exp}
&&g=\hbar\sqrt{\frac{4\pi a_{f} \Delta B \Delta\mu_{\rm mag}}{m_f}}\left(\frac{m_f\omega}{2\pi\hbar}\right)^{3/4} \, , \\
\label{delta_exp}
&&\delta=\Delta\mu_{\rm mag}(B-B_0) \, .
\end{eqnarray}
%%%%%%%%%%%%%%%%%%%%%%%%%%%%%%%%%%%%%%%%%%%%%%%%%%%%%%%%%%%%%%%%%%%%%%%%%%%%%%%%%%%%%%
Here $E_{r}^{f(b)}=h^2/2\lambda^2 m_{f(b)}$ is the recoil energy, $V_0$ is the amplitude of the optical lattice potential and $\lambda$ is the laser wavelength.  $a_f$, $a_b$,  and $a_{fb}$ are the fermion-fermion, boson-boson, and fermion-boson background scattering lengths. In our calculation we approximate the background boson-boson and the  Bose-Fermi scattering lengths by $a_b = 0.6 a_{f}$ \cite{Petrov04} and  $a_{fb} = 1.2 a_{f}$ \cite{Petrov05}. Furthermore, $B$ is the magnetic field, and $B_0$ and $\Delta B$ are the position of the Feshbach resonance and its width, respectively. $\Delta\mu_{\rm mag}$ is the difference in the magnetic moment between the closed and open channel of the Feshbach resonance. Finally, $m_f$ and $m_b$ are the respective masses of the fermions and bosons.

The Hamiltonian (\ref{Hamiltonian_full}) can be simplified  by the following rescaling:
\begin{eqnarray}
\label{mu}
\bar\mu=\mu-\frac{3\hbar\omega}{2} \, , \\
\bar\delta=\delta-\frac{3\hbar\omega}{2} \, ,
\end{eqnarray}
such that the factor $\frac{3\hbar\omega}{2}$ disappears.
%After this rescaling the factor $\frac{3\hbar\omega}{2}$ disappears.

\section{Method}\label{Method}

\subsection{Derivation of the effective single-band Hamiltonian }

The multi-band Hamiltonian derived so far is very complicated, since it both involves strong correlations and many bands. Simply neglecting the higher bands would lead to an incorrect description close to the Feshbach resonance, since the Feshbach parameter $g$ is very large and even exceeds the band gap \cite{Koetsier}. However, the filling of fermions in the higher bands is strongly suppressed by the band gap. This allows us to perform a mean-field decoupling in the higher bands \cite{Koetsier}. The lowest band is left untouched in this procedure, since the fermionic filling can be large there.

We thus perform the following decoupling for $l>0$ on each site:
\begin{equation}
\label{decoupling}
\hat {\cal H}_{fb}^{li}= g_l \left( \langle \hat b_i^\dagger \rangle  \hat c_{i\uparrow,l}^{\phantom\dagger}\hat c_{i\downarrow,l}^{\phantom\dagger}+\hat b_i^\dagger \langle  \hat c_{i\uparrow,l}^{\phantom\dagger}\hat c_{i\downarrow,l}^{\phantom\dagger} \rangle + h.c \right) \, .
\end{equation}
This step implies that the lowest band and the higher bands are only coupled in a mean-field way. They can thus be diagonalized separately, but are coupled by the mean-field self-consistency relations. The full Hamiltonian is now given by 
\begin{equation}
\label{Hamiltonian_SB}
\hat  {\cal H}=\hat {\cal H}_f^0+\hat {\cal H}_b+\hat {\cal H}_{fb}^0+\sum_{i}\hat {\cal H}_{b}^{\prime}(i) +\sum_{l,i}\hat {\cal H}_{fb}^{li}\, ,
\end{equation}
where the following terms are added to the bosonic part of the lowest band Hamiltonian:
\begin{equation}
\label{Hamiltonian_b_add}
\hat {\cal H}_{b}^{\prime}(i)= \sum_{l=1} g_l\left(\hat b_i^\dagger \langle \hat c_{i\uparrow,l}^{\phantom\dagger}\hat c_{i\downarrow,l}^{\phantom\dagger}\rangle + h.c\right)
=-\left(\Delta \hat b_i^\dagger + h.c.\right) \, .
\end{equation}
where the mean-field $\Delta$ has been defined as $\Delta=-\sum_{l=1}^\infty g_l \langle  \hat c_{i\uparrow,l}^{\phantom\dagger} \hat c_{i\downarrow,l}^{\phantom\dagger}\rangle$.
For each of the higher bands $l > 0$ we obtain the following Hamiltonian (here we suppress the site index $i$):
\begin{equation}
\label{Higher_bands}
\hat {\cal H}^l_{f}=\left( 
\begin{array}{c} 
\hat c_{l\uparrow}^\dagger \\ 
\hat c_{l\downarrow} 
\end{array} 
\right)\left( 
\begin{array}{cc}
 2l  \hbar\omega -\bar \mu & -g_l \langle \hat b \rangle \\ 
-g_l \langle \hat b^\dagger \rangle & -(2l  \hbar\omega -\bar \mu) 
\end{array} 
\right)\left( 
\begin{array}{c} 
\hat c_{l\uparrow} \\ 
\hat  c_{l\downarrow}^\dagger 
\end{array} 
\right) \, .
\end{equation}
The system described by Eqs. (\ref{Hamiltonian_SB}), (\ref{Hamiltonian_b_add}) and (\ref{Higher_bands}) needs to be solved self-consistently with respect to the mean fields 
$\langle \hat b \rangle$ and $\Delta$.

To diagonalize the Hamiltonian (\ref{Higher_bands}), one has to perform the following Bogoliubov transformation:
\begin{eqnarray}
\hspace{-2cm}\left(
\begin{array}{cc}
u_l & -v_l\\
v_l & u_l 
\end{array}
\right)\left( 
\begin{array}{cc} 
2l  \hbar\omega -\bar \mu & -g_l \langle \hat b \rangle \\ 
-g_l \langle \hat b^\dagger \rangle & -(2l  \hbar\omega -\bar \mu) 
\end{array} 
\right)\left(
\begin{array}{cc}
u_l & -v_l\\
v_l & u_l 
\end{array}
\right)^{-1}=\left(
\begin{array}{cc}
-\omega_l & 0\\
0 & w_l 
\end{array}
\right) \, ,
\end{eqnarray}
where
\begin{eqnarray}
\omega_l &=& \sqrt{(2l  \hbar\omega -\bar \mu)^2 + g_l^2 |\langle \hat b \rangle|^2} \, , \\
 u_l^2 &=& \frac{1}{2} + \frac{ 2l  \hbar\omega - \bar \mu}{2 \omega_l} \, , \\
v_l^2 &=& \frac{1}{2} - \frac{ 2l  \hbar\omega - \bar \mu}{2 \omega_l} \, , \\
u_l v_l &=& \frac{g_l \langle \hat b \rangle}{2 \omega_l} \, . 
\end{eqnarray}
This leads to the following expectation values:
\begin{eqnarray}
&&n_{l}^F = 2 v_l^2+2(u_l^2-v_l^2)f(\omega_l) %\nonumber\\
=1 - \frac{ 2l  \hbar\omega - \bar \mu}{\omega_l}\tanh\left(\frac{\omega_l}{2kT}\right) \, , \\
&&|\langle \hat c_{l\uparrow} \hat c_{l\downarrow} \rangle |=| u_l v_l | \tanh\left(\frac{\omega_l}{2kT}\right) %\nonumber\\
= \left| \frac{g_l \langle \hat b \rangle}{2 \omega_l} \right| \tanh\left(\frac{\omega_l}{2kT}\right) \, ,
\end{eqnarray}
where $f(\omega_l)$ is the Fermi function and $T$ is the temperature. We use absolute values in the equation for 
$ \langle \hat c_{l\uparrow} \hat c_{l\downarrow} \rangle$ because of the ambiguity of the sign, arising from 
the fact that still a divergence has to be subtracted (see below).

The total number of fermions is equal to:
\begin{equation}
 n_{{\rm tot}}^F = n_{0}^F + \sum_{l=1}^{\infty}\left( 1 - \frac{ 2l  \hbar\omega - \bar \mu}{\omega_l}\tanh\left(\frac{\omega_l}{2kT}\right)\right) \, .
\end{equation}
This is a converging sum, which can be evaluated numerically.

From Eq. (\ref{Hamiltonian_b_add}) follows that we have to evaluate the sum
\begin{equation}
\sum_{l=1} g_l \langle \hat c_{l\uparrow} \hat c_{l\downarrow} \rangle = 
%\pm \sum_{l=1} g_l \frac{g_l \langle b \rangle}{2 \omega_l} = 
\pm \langle \hat  b \rangle \sum_{l=1} \frac{g_l^2}{2 \omega_l}\tanh\left(\frac{\omega_l}{2kT}\right) \, .
\end{equation}
which is divergent. This divergence always arises in the gap equation of the BCS model when the T-matrix is approximated by a delta-potential \cite{Koetsier, Busch}. One way to solve this problem is by using a pseudo-potential instead of the delta-potential \cite{Busch}. Here, however, we follow Ref. \cite{Koetsier} and explicitly isolate the diverging contribution from the sum.   

First, we notice that for large $l$, $\omega_l$ can be approximated by $\omega_l = 2l  \hbar\omega -\bar \mu$ and 
$\tanh\left(\frac{\omega_l}{2kT}\right)\simeq 1$. Therefore
\begin{eqnarray}
&&\hspace{-1.5cm}\sum_{l=1} \frac{g_l^2}{2 \omega_l} \tanh\left(\frac{\omega_l}{2kT}\right) %\nonumber\\
\simeq \left(\sum_{l=1}^N \frac{g_l^2}{2 \omega_l}\tanh\left(\frac{\omega_l}{2kT}\right) +
\sum_{l=N+1}^\infty\frac{g_l^2}{2(2l  \hbar\omega -\bar \mu)}\right)  \nonumber\\
&&\hspace{-0.5cm}=\left(\sum_{l=1}^N \frac{g_l^2}{2 \omega_l}\tanh\left(\frac{\omega_l}{2kT}\right)-
\sum_{l=0}^N\frac{g_l^2}{2(2l  \hbar\omega -\bar \mu)} \right. %\nonumber\\
\left.+\sum_{l=0}^\infty\frac{g_l^2}{2(2l  \hbar\omega -\bar \mu)}\right) \, . 
\label{sum_1}
\end{eqnarray}
Here $N$ is a large integer number (in our calculation we took $N=500$).

The first two terms of Eq. (\ref{sum_1}) are finite sums, but the last term diverges. This sum is known from the literature \cite{Busch,Abramowitz}.  To separate the diverging part, we have to take the following limit:
\begin{eqnarray}
\label{limit}
&&\hspace{-1cm}\sum_{l=0}^\infty\frac{g_l^2}{2(2l  \hbar\omega -\bar \mu)}=
\sum_{l=0}^\infty\frac{g_0^2 L^{(1/2)}_{l}(0) }{2(2l \hbar\omega -\bar \mu)}=
\lim_{r\rightarrow 0}\sum_{l=0}^\infty\frac{g_0^2 L^{(1/2)}_{l}(r) }{2(2l \hbar\omega -\bar \mu)}  \nonumber \\
&&\hspace{0.45cm}=-\lim_{r\rightarrow 0}\left(\frac{g_0^2 \sqrt{\pi} \Gamma( -\bar \mu/2\hbar \omega)/\Gamma( -\bar \mu/2\hbar \omega- 1/2)}{2\hbar \omega} %\right.\nonumber\\ 
- \frac{\sqrt{\pi}}{r}+{\cal O}(r) \right) \, .
\end{eqnarray}
Since the diverging part $\sqrt{\pi}/r$ is independent of the model parameters, we can cure the divergence by neglecting this term \cite{Koetsier, Busch}. Doing so, we obtain
\begin{eqnarray}
\label{Delta}
\Delta &=& -\sum_{l=1}^{\infty} g_l \langle \hat c_{l\uparrow} \hat c_{l\downarrow}\rangle %\nonumber\\
= \pm \langle \hat b \rangle 
\left( \frac{g_0^2 \sqrt{\pi} \Gamma( -\bar \mu/\hbar \omega)/\Gamma( -\bar \mu/\hbar \omega- 1/2)}{2\hbar \omega} \right. \nonumber\\
&+&\left. \sum_{l=0}^N \frac{g_l^2}{2(2l  \hbar\omega -\bar \mu)} 
- \sum_{l=1}^N \frac{g_l^2}{2 \omega_l}\tanh\left(\frac{\omega_l}{2kT}\right)\right) \, .
\end{eqnarray}
We now fix the sign by requiring $\Delta>0$, since this solution minimizes the (free) energy.

Summarizing, we have reduced the multi-band problem to an effective single-band Hamiltonian:
\begin{equation}
\label{end_hamiltonian}
\hat {\cal H}=\hat {\cal H}_f^0+ \hat {\cal H}_{b} + \hat {\cal H}'_{b}+\hat {\cal H}_{fb}^0  \, ,
\end{equation}
where $\hat {\cal H}_f^0$, $\hat {\cal H}_{b}$,  $\hat {\cal H}_{fb}^0$ and $\hat {\cal H}^\prime_{b}$ are given by Eqs. (\ref{Hamiltonian_f0}), (\ref{Hamiltonian_b}), (\ref{Hamiltonian_fb0}) and  (\ref{Hamiltonian_b_add}), respectively.

The chemical potential $\mu$ has to be adjusted such that the total filling is equal to the desired value $n_{\rm tot}$:
\begin{equation}
 2n^b_0 + n_0^F + \sum_{l=1}^{\infty} \left( 1 - \frac{ 2l  \hbar\omega - \bar \mu}{\omega_l} \tanh\left(\frac{\omega_l}{2kT}\right) \right) = n_{\rm tot} \, . 
\end{equation}

This leads to the following self-consistency loop: we start from an initial guess of the superfluid order parameter $\langle \hat b \rangle$ and calculate $\Delta$ using Eq. (\ref{Delta}). As a result we know all parameters in the Hamiltonian (\ref{end_hamiltonian}), and can find its eigenvalues and eigenvectors, and correspondingly calculate new correlation functions, including the superfluid order parameter $\langle \hat b \rangle$. With this step  the self-consistency loop is closed.

It is worth noting that the effective single-band Hamiltonian we have derived here is different from the effective single-band model in terms of dressed particles derived in other approaches \cite{Dickerscheid05b, Gubbels06}: the bosons and fermions in our Hamiltonian correspond to the bare particles in the lowest band.

\subsection{Generalized Dynamical Mean-Field Theory}

To analyze the Hamiltonian (\ref{end_hamiltonian}) we use Generalize Dynamical Mean Field Theory (GDMFT) which is explained in detail in Ref. \cite{Titvinidze08, Titvinidze09}. Here we only mention that within GDMFT one considers a single site which is self-consistently coupled to a dynamical fermionic bath  corresponding to DMFT \cite{DMFT1,DMFT2}  and a static bosonic mean-field corresponding to bosonic Gutzwiller \cite{Gutzwiller1,Gutzwiller2,Gutzwiller3}.  These are the leading order contributions in a $1/z$-expansion of the effective action ($z$ being the lattice coordination number). Hence GMDFT is exact in infinite dimensions and expected to be a good approximation for the cubic lattice considered here, for which $z=6$. The typical accuracy for low-temperature expectation values is around 20 percent.

In the specific case considered here the system is described by a generalized single impurity Anderson model (GSIAM) with the following Anderson Hamiltonian:
%%%%%%%%%%%%%%%%%%%%%%%%%%%%%%%%%%%%%%%%%%%%%%%%%%%%%%%%%%%%%%%%%%%%%%%%%%%%%%%%%%%%%%
\begin{eqnarray}
\label{GSIAM}
&&\hspace{-2.5cm}\hat{\mathcal H}^{\rm And}=
\hat{\mathcal H}_{f}^{\rm And} + \hat{\mathcal H}_{fb}^{\rm And} +\hat{\mathcal H}_{b}^{\rm And} \, , \\
&&\hspace{-2.5cm}\hat{\mathcal H}_{b}^{\rm And}=- \left[(z t_b \varphi^{\phantom *}+\Delta) \hat b^\dagger + h.c.\right]
+ \frac{U_{b}}{2} \hat n^{b}(\hat n^{b} -1) - (2\bar\mu-\bar\delta)\hat n^b \, , \nonumber  \\
&&\hspace{-2.5cm}\hat{\mathcal H}_{fb}^{\rm And} = U_{fb} \hat n^f  \hat n^{b}+ 
g_0 \left(\hat b_i^\dagger \hat c_{\uparrow}^{\phantom\dagger}\hat c_{\downarrow}^{\phantom\dagger}+h.c\right) \, , \nonumber  \\
&&\hspace{-2.5cm}\hat{\mathcal H}_{f}^{\rm And} = - \bar\mu \hat n^{f} + U_{f}\hat n^{f}_{\uparrow}\hat n^{f}_{\downarrow}
+\sum_{k,\sigma}\Bigl\{\varepsilon_{k} \hat a^{\dagger}_{k\sigma}\hat a_{k\sigma}^{\phantom\dagger} 
+V_{k}\left(\hat c^{\dagger}_{\sigma}\hat a_{k \sigma}^{\phantom\dagger}+h.c.\right)\Bigr\}
+ \sum_{k} W_{k}\left(\hat a^{\dagger}_{k\uparrow}\hat a_{k\downarrow}^{\dagger}+h.c.\right)  \, , \nonumber 
%&&\hspace{-2.5cm}\hat{\mathcal H}_{f} = - \mu_{\sigma f} \hat n^{f} + U_{f}\hat n^{f}_{\uparrow}\hat n^{f}_{\downarrow}
%+\sum_{k,\sigma}\Bigl\{\varepsilon_{k\sigma} \hat a^{\dagger}_{k\sigma}\hat a_{k\sigma}^{\phantom\dagger} 
%+V_{k\sigma}\left(\hat c^{\dagger}_{\sigma}\hat a_{k \sigma}^{\phantom\dagger}+h.c.\right)\Bigr\}
%+ \sum_{k} W_{k}\left(\hat a^{\dagger}_{l\uparrow}\hat a_{k\downarrow}^{\dagger}+h.c.\right)  \, , \nonumber  
\end{eqnarray}
%%%%%%%%%%%%%%%%%%%%%%%%%%%%%%%%%%%%%%%%%%%%%%%%%%%%%%%%%%%%%%%%%%%%%%%%%%%%%%%%%%%%%%
where $z$ is the lattice coordination number and $\varphi= \langle \hat b \rangle$ is the superfluid order parameter. $k$ labels the noninteracting orbitals of the effective bath, $V_{k}$ are the corresponding fermionic hybridization matrix elements, $W_k$ describes the superfluid properties of the bath and $a_{k\sigma}^\dagger$ is the creation operator of a fermion in the $k^{\rm th}$ orbital of the bath with spin $\sigma$.
$\hat n_{f\sigma}=\hat c_{\sigma}^\dagger c_{\sigma}$ is the fermionic number operator and $\hat n=\hat n_{f\uparrow}+\hat n_{f\downarrow}$.

To solve the Anderson Hamiltonian we use exact diagonalization as impurity solver \cite{DMFT_review,Caffarel,Si,Toschi}. 
In this algorithm the infinite number of orbitals in the Hamiltonian (\ref{GSIAM}) is truncated and only a finite number of $n_s$ orbitals is considered. The resulting finite-size problem is fundamentally different from the finite-size problem of a finite number of lattice sites of the original Hubbard model, and the truncation procedure can be viewed as using a finite number of parameters (energy scales) to describe the local dynamics encoded in the Weiss Green's function:
%%%%%%%%%%%%%%%%%%%%%%%%%%%%%%%%%%%%%%%%%%%%%%%%%%%
\begin{eqnarray}
\label{Weiss_Green_sigma}
&&{\cal G}^{-1}_{\rm And}(i\omega_{n})={\cal G}^{-1}_{\sigma,{\rm And}}(i\omega_{n})=
i\omega_{n}+\bar \mu +\sum_{l=1}^{n_s}V_{l\sigma}^2 
\frac{i\omega_n+\varepsilon_{l}}{\varepsilon_{l}^2+\omega_n^2+W_l^2 } 
%i\omega_{n}\mu_\sigma +\sum_{l=1}^{n_s}V_{l\sigma}^2 
%\frac{i\omega_n+\varepsilon_{l\bar\sigma}}{(\varepsilon_{l\sigma}-i\omega_n)(\varepsilon_{l\bar\sigma}+i\omega_n)+W_l^2 } 
\, ,\\
\label{Weiss_Green_SC}
&&{\cal F}^{-1}_{\rm And}(i\omega_{n})=\sum_{l=1}^{n_s}
\frac{V_{l}^2 W_l}{\varepsilon_l^2+\omega_n^2+W_l^2 } \, ,
%\frac{V_{l\uparrow}V_{l\downarrow}W_l}{(\varepsilon_{l\sigma}-i\omega_n)(\varepsilon_{l\bar\sigma}+i\omega_n)+W_l^2 } \, ,
\end{eqnarray}
%%%%%%%%%%%%%%%%%%%%%%%%%%%%%%%%%%%%%%%%%%%%%%%%%%%
where  $\beta$ is the inverse temperature and $\omega_n=(2n+1)\pi/\beta$ are the Matsubara frequencies. 

To close the self-consistency loop by using the lattice Dyson equation we calculate the normal and superfluid Green's functions which can be written as follows  \cite{DMFT_review}
%%%%%%%%%%%%%%%%%%%%%%%%%%%%%%%%%%%%%%%%%%%%%%%%%%%
\begin{eqnarray}
\label{G_Green_int}
&&\hspace{-0.75cm}G(i\omega_n)=G_\sigma(i\omega_n)=\int_{-\infty}^{\infty} d\varepsilon D(\varepsilon) \frac{\zeta^*-\varepsilon}{|\zeta-\varepsilon|^2+\Sigma_{SC}^2} \, ,\\ 
\label{F_Green_int}
&&\hspace{-0.75cm}F(i\omega_n)=-\Sigma_{SC}(i\omega_n)\int_{-\infty}^{\infty} d\varepsilon D(\varepsilon) \frac{1}{|\zeta-\varepsilon|^2+\Sigma_{SC}^2}\, ,
\end{eqnarray}
%%%%%%%%%%%%%%%%%%%%%%%%%%%%%%%%%%%%%%%%%%%%%%%%%%%
where $\zeta=i\omega_n+\bar \mu-\Sigma(i\omega_n)$ and  $D(\varepsilon)$ is the non-interacting density of the states of the cubic lattice.  $\Sigma(i\omega_n)$ and $\Sigma_{SC}(i\omega_n)$ are the normal and superfluid self-energies, which as shown in \ref{Appendix} can be  expressed via a set of higher order Green's functions:
%%%%%%%%%%%%%%%%%%%%%%%%%%%%%% 
\begin{eqnarray}
\label{SE_sigma}
&&\hspace{-1cm}\Sigma(i\omega_n)=\Sigma_\sigma(i\omega_n)=\frac{\left(U_f Q_{ff\sigma}(i\omega_n)+U_{fb} Q_{fb\sigma}(i\omega_n)+\sigma g Q_{g\bar\sigma\sigma}^{*}(i\omega_n)\right) G_{\bar\sigma}^*(i\omega_n)}{G_\sigma(i\omega_n)G_{\bar\sigma}^*(i\omega_n)+F(\sigma i\omega_n)F^{*}(\bar\sigma i\omega_n)}\nonumber\\
&&\hspace{0.5cm}+\frac{\left(\sigma U_f Q_{ff,\sigma\bar\sigma}(i\omega_n)+\sigma U_{fb} Q_{fb\sigma\bar\sigma}(i\omega_n)+g Q_{g\bar\sigma}^*(i\omega_n) \right) F^{*}(\bar\sigma i\omega_n)}{ G_\sigma(i\omega_n)G_{\bar\sigma}^*(i\omega_n)+F(\sigma i\omega_n)F^{*}(\bar\sigma i\omega_n) } \, , \\
\label{SE_SC}
&&\hspace{-1cm}\Sigma_{SC}(i\omega_n)=\frac{\left(U_f Q_{ff\uparrow}(i\omega_n)+U_{fb} Q_{fb\uparrow}(i\omega_n)+g Q_{g\downarrow\uparrow}^{*}(i\omega_n)\right) 
F(i\omega_n)}{G_\uparrow(i\omega_n)G_\downarrow^*(i\omega_n)+F(i\omega_n)F^{*}(-i\omega_n) }\nonumber\\
&&\hspace{0.5cm}-\frac{\left(U_f Q_{ff,\uparrow\downarrow}(i\omega_n)+U_{fb} Q_{fb\uparrow\downarrow}(i\omega_n)+g Q_{g\downarrow}^*(i\omega_n)\right) G_{\uparrow}(i\omega_n)}{G_\uparrow(i\omega_n)G_\downarrow^*(i\omega_n)+F(i\omega_n)F^{*}(-i\omega_n) }  \, . 
%\label{SE_SC_star}
%&&\Sigma_{SC}^*(i\omega_n)=\frac{\left(U_f Q_{ff\downarrow}^*(i\omega_n)+U_{fb} Q_{fb\downarrow}^*(i\omega_n)-g Q_{g\uparrow\downarrow}(i\omega_n) \right) F^*(-i\omega_n)}{G_\uparrow(i\omega_n)G_\downarrow^*(i\omega_n)+F(i\omega_n)F^{*}(-i\omega_n) } \\
%&&\hspace{1.5cm}+\frac{\left(U_f Q_{ff,\downarrow\uparrow}^{*}(i\omega_n)+U_{fb} Q_{fb\downarrow\uparrow}^{*}(i\omega_n)-g Q_{g\uparrow}(i\omega_n) \right) G_\downarrow^*(i\omega_n)}{G_\uparrow(i\omega_n)G_\downarrow^*(i\omega_n)+F(i\omega_n)F^{*}(-i\omega_n) }\nonumber
\end{eqnarray}
%%%%%%%%%%%%%%%%%%%%%%%%%%%%%%
Here $G(i\omega_n)=\langle\langle c_{\sigma,0}^{\phantom\dagger}, c_{\sigma,0}^{\dagger} \rangle\rangle_\omega$ and $F(i\omega_n)=\langle\langle c_{\uparrow,0}^{\phantom\dagger}, c_{\downarrow,0}^{\phantom\dagger} \rangle\rangle_\omega$ are the normal and superfluid single particle Green's functions. In addition we have also defined the following additional interacting Green's functions:
$Q_{ff\sigma}(i\omega_n)=\langle\langle \hat f^{\phantom\dagger}_\sigma \hat f^{\dagger}_{\bar\sigma} \hat f^{\phantom\dagger}_{\bar\sigma} ,\hat f^{\dagger}_\sigma \rangle\rangle_\omega$, 
$Q_{ff\sigma\bar\sigma}(i\omega_n)=\langle\langle \hat f^{\phantom\dagger}_\sigma \hat f^{\dagger}_{\bar\sigma} \hat f^{\phantom\dagger}_{\bar\sigma} , \hat f^{\phantom\dagger}_{\bar\sigma} \rangle\rangle_\omega $, 
$Q_{fb\sigma}(i\omega_n)=\langle\langle \hat f^{\phantom\dagger}_\sigma \hat b^{\dagger}\hat b, \hat f^{\dagger}_\sigma \rangle\rangle_\omega $, $Q_{fb\sigma\bar\sigma}(i\omega_n)=\langle\langle \hat f^{\phantom\dagger}_\sigma \hat  b^{\dagger}\hat b, \hat f^{\phantom\dagger}_{\bar\sigma} \rangle\rangle_\omega $, 
$Q_{g\sigma}(i\omega_n)=\langle\langle \hat f^{\phantom\dagger}_\sigma \hat b^{\dagger}, \hat f^{\dagger}_\sigma \rangle\rangle_\omega $ and $Q_{g\sigma\bar\sigma}(i\omega_n)=\langle\langle \hat f^{\phantom\dagger}_\sigma \hat b^{\dagger}, \hat f^{\phantom\dagger}_{\bar\sigma} \rangle\rangle_\omega$. Here
%%%%%%%%%%%%%%%%%%%%%%%%%%%%%%%%%%%%%%%%%%%%%%%%%%%
\begin{equation}
\label{Green}
\langle\langle \hat A,\hat B\rangle\rangle_\omega=
-\frac{1}{Z}\sum_{n,m}\langle n |\hat A |m \rangle\langle m| \hat B |n\rangle\frac{e^{-\beta E_n}+e^{-\beta E_m}}{E_m-E_n-i\omega_n} \, ,
\end{equation}
%%%%%%%%%%%%%%%%%%%%%%%%%%%%%%%%%%%%%%%%%%%%%%%%%%%
and 
%%%%%%%%%%%%%%%%%%%%%%%%%%%%%%%%%%%%%%%%%%%%%%%%%%%
\begin{equation}
\label{stat_sum_method}
Z=\sum_{n} e^{-\beta E_n}
\end{equation}
%%%%%%%%%%%%%%%%%%%%%%%%%%%%%%%%%%%%%%%%%%%%%%%%%%%
is the partition function.

The relation between the Weiss field and the Green's function is given by the local Dyson equation:
%%%%%%%%%%%%%%%%%%%%%%%%%%%%%%%%%%%%%%%%%%%%%%%%%%%
\begin{equation}
\label{Dyson_SC}
{\hat{\cal G}}^{-1}_{\rm ex}(i\omega_n)={\hat\Sigma}(i\omega_n)+{\hat G}^{-1}(i\omega_n)  \, ,
\end{equation}
%%%%%%%%%%%%%%%%%%%%%%%%%%%%%%%%%%%%%%%%%%%%%%%%%%%
where 
%%%%%%%%%%%%%%%%%%%%%%%%%%%%%%%%%%%%%%%%%%%%%%%%%%%
\begin{equation}
\label{Green_matrix}
\hat G(i\omega_n)= \left(
\begin{array}{cc}
G(i\omega_n)& F(i\omega_n)\\
F(i\omega_n) & -G^*(i\omega_n)
\end{array}
\right)
\end{equation}
%%%%%%%%%%%%%%%%%%%%%%%%%%%%%%%%%%%%%%%%%%%%%%%%%%%
is the matrix of interacting Green's functions, 
%%%%%%%%%%%%%%%%%%%%%%%%%%%%%%%%%%%%%%%%%%%%%%%%%%%
\begin{equation}
\label{Self_energy_matrix}
\hat \Sigma(i\omega_n)= \left(
\begin{array}{cc}
\Sigma(i\omega_n)& \Sigma_{SC}(i\omega_n)\\
\Sigma_{SC}(i\omega_n) & -\Sigma^*(i\omega_n)
\end{array}
\right)
\end{equation}
%%%%%%%%%%%%%%%%%%%%%%%%%%%%%%%%%%%%%%%%%%%%%%%%%%%
is the self-energy matrix  and 
%%%%%%%%%%%%%%%%%%%%%%%%%%%%%%%%%%%%%%%%%%%%%%%%%%%
\begin{equation}
\label{Weiss_Green_matrix}
\hat {\cal G}_{\rm ex}(i\omega_n)= \left(
\begin{array}{cc}
{\cal G}_{\rm ex}(i\omega_n)& {\cal F}_{\rm ex}(i\omega_n)\\
{\cal F}_{\rm ex}(i\omega_n) & -{\cal G}_{\rm ex}^*(i\omega_n)
\end{array}
\right)
\end{equation}
%%%%%%%%%%%%%%%%%%%%%%%%%%%%%%%%%%%%%%%%%%%%%%%%%%%
is the matrix of Weiss Green's functions. 

To determine new parameters for the Anderson Hamiltonian, we fit the Weiss functions calculated from  (\ref{Weiss_Green_sigma}) and (\ref{Weiss_Green_SC}) to the ones calculated from the eigenstates of the Anderson Hamiltonian via the local Dyson equation (\ref{Dyson_SC}). We use a steepest decent method with the following norm:
%%%%%%%%%%%%%%%%%%%%%%%%%%%%%%%%%%%%%%%%%%%%%%%%%%%
\begin{eqnarray}
\label{feet}
\hspace{-2.5cm}\chi=\frac{1}{2(N_{\rm max}+1)}\sum_{n=0}^{N_{max}}\frac{1}{2n+1}
\left(|{\cal G}^{-1}_{\rm And}(i\omega_{n})-{\cal G}^{-1}_{\rm ex}(i\omega_{n})|^2 
+ |{\cal F}^{-1}_{\rm And}(i\omega_{n})-{\cal F}^{-1}_{\rm ex}(i\omega_{n})|^2\right) \, , \nonumber\\
\end{eqnarray}
%%%%%%%%%%%%%%%%%%%%%%%%%%%%%%%%%%%%%%%%%%%%%%%%%%%
where $N_{\rm max}$ is the number of Matsubara frequencies taken into account.

The minimization procedure works as follows: we start from an initial guess of the GSIAM parameters ($\epsilon_{l\sigma}$, $V_{l\sigma}$ and $W_l$), and then knowing the local Green's functions calculated from Eq. (\ref{Green}) and the self-energies calculated from (\ref{SE_sigma}) and (\ref{SE_SC}) we calculate the lattice Green's function according to Eqs. (\ref{G_Green_int}) and (\ref{F_Green_int}). Subsequently, using the Dyson equation (\ref{Dyson_SC}) we can calculate the Weiss Green's functions ${\cal G}^{-1}_{\sigma,ex}(i\omega_{n})$ and ${\cal F}^{-1}_{\sigma,ex}(i\omega_{n})$. The next step is to fit this result by the parameterization in Eqs. (\ref{Weiss_Green_sigma}) and (\ref{Weiss_Green_SC}) and thus to find a new set of parameters for the GSIAM. These new parameters serve as input for the next iteration. This procedure is repeated until convergence is reached.

\subsection{Calculation of the critical temperature}

The combination of the mean-field approximation in the higher bands and GDMFT explained so far, however, leads to a problem. Both approximations involve the superfluid order parameter $\langle \hat b_i \rangle$. The mean-field approximation for the higher bands implies that the local correlator $\langle \hat b_i^\dagger \hat c_{i\uparrow,l}^{\phantom\dagger} \hat c_{i\downarrow,l}^{\phantom\dagger} \rangle$ is approximated by $\langle \hat b_i^\dagger \rangle \langle \hat c_{i\uparrow,l}^{\phantom\dagger}\hat c_{i\downarrow,l}^{\phantom\dagger} \rangle$. The GDMFT scheme, on the other hand, involves the approximation to replace the non-local correlator $\langle \hat b_i^\dagger \hat b_j \rangle$ by $\langle \hat b_i^\dagger \rangle \langle \hat b_j \rangle$. This means that $\langle \hat b \rangle$ both measures the local phase coherence between bosons and fermions and the non-local bosonic long range order. However, these are two very different quantities which generally cannot be described by a single mean-field order parameter. At zero temperature, this problem is not too severe, because in this case one expects both long-range order and on-site boson-fermion coherence, so that $\langle \hat b \rangle$ is large for both reasons. At finite temperature, however, this becomes a real problem, because the bosonic long range order is expected to vanish at temperatures of the order of the bosonic hopping $t_b$. The local boson-fermion coherence, on the other hand, persists for much higher temperatures, since the coupling $g$ is orders of magnitudes larger than $t_b$. Indeed, we find that in the approximation outlined above the full  GDMFT calculations are in good agreement with a single site approximation.  In the single-site approach the impurity site couples neither to the fermionic nor to the bosonic bath and long range order cannot be inferred. The critical temperature obtained from this calculation can be identified with the pair breaking temperature $T_{\rm pair}$, which is much higher than the relevant temperature in experiments. 

The scheme explained in the previous subsection can therefore not be used to infer the critical temperature for superfluid long range order. In order to do so, we have to modify the approximation and remove the ambiguous nature of  the order parameter $\langle \hat b \rangle$. This is made possible by the observation that the term  $\Delta \hat b^\dagger$ in the Hamiltonian merely renormalizes the self-energy of the bosons: in the BEC regime the bosons are in a coherent state and this term is equivalent to a shift of the bosonic chemical potential. This is also clear from the treatment in \cite{Koetsier}, where terms from higher bands enter the bosonic self-energy. To make this more explicit, we write
\begin{eqnarray}
\label{Delta2}
\Delta &=& -\sum_{l=1}^{\infty} g_l \langle \hat c_{l\uparrow} \hat c_{l\downarrow} \rangle  
= \pm \langle \hat  b \rangle \left( \frac{g_0^2 \sqrt{\pi} \Gamma( -\bar \mu/\hbar \omega)/\Gamma( -\bar \mu/\hbar \omega- 1/2)}{2\hbar \omega} \right. \nonumber\\
&+&\left. \sum_{l=0}^N \frac{g_l^2}{2(2l  \hbar\omega -\bar \mu)} 
-\sum_{l=1}^N \frac{g_l^2}{2 \omega_l}\tanh\left(\frac{\omega_l}{2kT}\right)\right) \equiv\langle \hat b \rangle \Delta' \, . 
\end{eqnarray}
We can therefore replace the term
\begin{equation}
- \left( \Delta \hat b^\dagger + h.c.\right) = - \left( \Delta' \langle \hat b \rangle  \hat b^\dagger + h.c. \right)
\end{equation}
in the Hamiltonian, by
\begin{equation}
- \Delta' \hat b^\dagger \hat b \, , 
\end{equation}
such that terms from the higher bands only renormalize the chemical potential. We remark here, that this might look like an additional approximation. However, one has to keep in mind that the term connecting $\Delta$ to the bosonic creation operator originated from the mean-field approximation in the higher bands. By treating the contribution of higher bands within the bosonic self-energy we are therefore restoring part of the mean-field approximation made in the previous step. Indeed, by using second order perturbation theory in the couplings to higher bands (which is justified if the band energy exceeds the Feshbach coupling), we can also obtain this correction directly as part of the bosonic self-energy, without invoking a mean-field decoupling.

This improved approximation for treating higher bands gives for $T = 0$ similar results as before; in particular the position of the transition to the Mott insulator is in good approximation the same. The superfluid order parameter is smaller, as expected. However, for nonzero temperatures this improved approximation scheme allows for a calculation of the critical temperature for superfluid long range order, which was not possible in the previous approximation.

%%%%%%%%%%%%%%%%%%%%%%%%%%%%%%%%%%%%%%%%%%%%%%%%%%%%%%%%%%%%%%%%%%%%%%%%%%%%%%%%%%%%%%%%%%%%%%%
\begin{figure}[hbpt]
\begin{center}
%\subfigure[]{
%\label{Scb1}
%\begin{minipage}[b]{0.45\textwidth}
%\centering \includegraphics[width=1.0\textwidth]{T0_1.jpg}
%\end{minipage}}
%\hspace{0.25cm}
%\subfigure[]{
%\label{nfnb1}
%\begin{minipage}[b]{0.45\textwidth}
%\centering \includegraphics[width=1.0\textwidth]{T0_2.jpg}
%\end{minipage}}\\
\subfigure[]{
\label{Scb2}
\begin{minipage}[b]{0.45\textwidth}
\centering \includegraphics[width=1.0\textwidth]{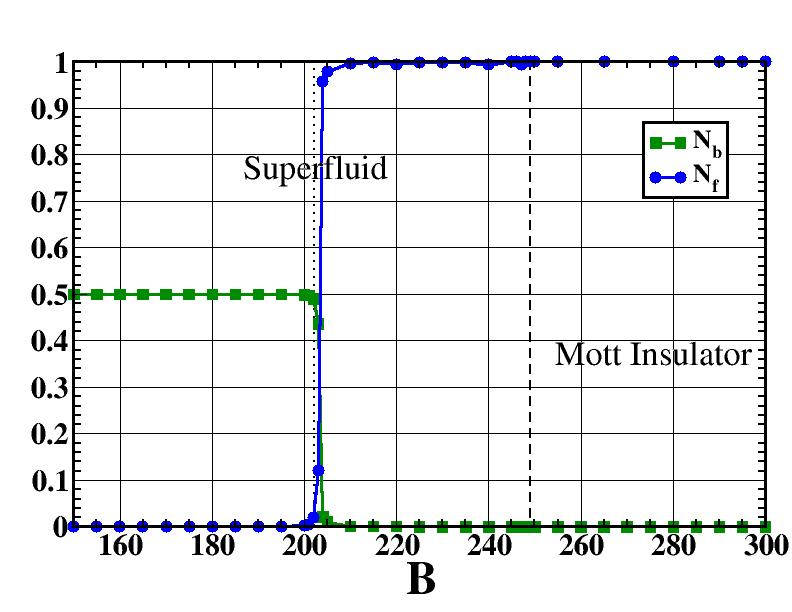}
\end{minipage}}
\hspace{0.25cm}
\subfigure[]{
\label{nfnb2}
\begin{minipage}[b]{0.45\textwidth}
\centering \includegraphics[width=1.0\textwidth]{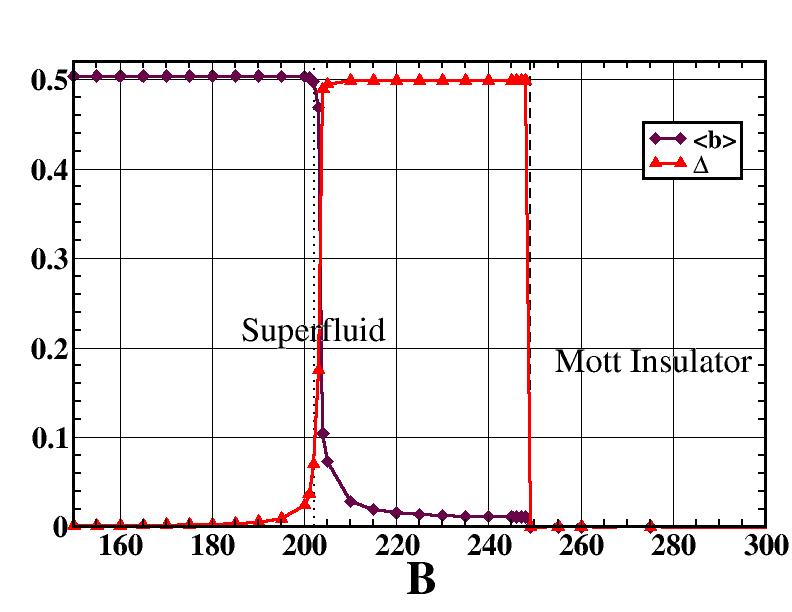}
\end{minipage}}\\
\caption{Bosonic and fermionic filling (a) and superfluid order parameters (b) as a function of magnetic field $B$ for $T=0$. The dotted line corresponds to the Feshbach resonance, while the dashed line indicates the phase transition from the superfluid phase into the Mott insulator phase. The magnetic field is measured in units of Gauss.}
%\caption{Plot of bosonic and fermionic filling and superfluid order parameters as a function of magnetic field $B$ for $T=0$. In sub-figure (a) and (b) we present results obtained by a mean-field decoupling for both local and non-local correlations, while in the sub-figures (c) and (d) only non-local correlations are decoupled (for details see the text). In the sub-figures (a) and (c) we show the filling of fermions and bosons and in the sub-figures (b) and (d) the superfluid order parameters. The dotted line corresponds to the Feshbach resonance, while the dashed line indicates the phase transition from the superfluid phase into the Mott insulator phase. The magnetic field is measured in units of Gauss.}
\label{T0}
\end{center}
\end{figure}
%%%%%%%%%%%%%%%%%%%%%%%%%%%%%%%%%%%%%%%%%%%%%%%%%%%%%%%%%%%%%%%%%%%%%%%%%%%%%%%%%%%%%%%%%%%%%%%%%%%%%%%%

\section{Results}\label{Results}

We study a mixture of potassium atoms ($^{40}{\rm K}$) and Feshbach molecules in a three-dimensional optical lattice. The on-site harmonic oscillator frequency is chosen to be $\omega=2\pi\times 58275 $Hz, which corresponds to a lattice with wavelength $\lambda=806$nm and Rabi frequency of $\Omega_R=2\pi\times1.43$GHz. The Feshbach resonance considered here is at $B=202.1$G and the width of the resonance is $7.8$ G \cite{Regal}.  The difference between the magnetic moments of the closed and open channels of the Feshbach resonance is $\Delta \mu = 16/9 \mu_B$, where $\mu_B$ is the Bohr magneton. The total filling per lattice site in our calculation is $n_{\rm tot}=1$. 

\subsection{Zero temperature}
First  we consider the case of zero temperature. Our calculations for the ground state are summarized in Fig. \ref{T0}. Deep in the BEC regime only bosonic molecules are present. When the magnetic field is increased, close to resonance the number of fermions is increasing and the number of bosons decreasing. Above the resonance we mainly have fermions and the number of bosons is small. Both fermions and bosons are superfluid. We remark again that here we describe the physics in terms of bare bosons and fermions: in terms of dressed particles as in Ref. \cite{Koetsier}, these are still molecular bosons and the BEC/BCS crossover takes place when the bosonic self-energy crosses twice the Fermi energy \cite{Koetsier}. However, in the case of half-filled fermions, this crossover is intercepted by a first order phase transition to a fermionic Mott insulator state, which happens at a critical value of the magnetic field of $B=249$G. Calculations which include only the lowest band of the Bose-Fermi-Hubbard model (as well as with one and two exited bands), yield this  transition into the Mott insulator phase already close to the Feshbach resonance at $B\simeq 205$G (results not shown). This implies that to capture the superfluid region $205\rm{T} \lesssim B < 249\rm{T}$ higher bands which renormalize the bosonic self-energy are crucial.

Another point worth noting is the first-order nature of the transition to the Mott-insulating state. In contrast, if one integrates out the bosonic degree of freedom and describes the Feshbach resonance in terms of an effective, attractive interaction between the fermions within a single channel model for the lowest band, one finds a different scenario. In this case the induced attractive interaction dominates, until it is cancelled by the repulsive background interaction. This means that within the single channel approximation one finds a regime with a normal Fermi-liquid phase in between the BEC/BCS phase and the Mott insulator, which is absent in our phase diagram.  This directly indicates that the effect of higher bands is crucial to capture the first-order transition between the superfluid and insulator phase.
%Another point worth noting is the first-order nature of the transition to the Mott-insulating state. This can only be captured if one explicitly keeps the bosonic molecular degree of freedom in the theory, as we did here. If one integrates out the bosonic degree of freedom and describes the Feshbach resonance in terms of an effective, attractive interaction between the fermions within a single channel model for the lowest band, one finds a different scenario. In this case the induced attractive interaction dominates, until it is cancelled by the repulsive background interaction. This means that within the single channel approximation one finds a regime with a normal Fermi-liquid phase in between the BEC/BCS phase and the Mott insulator, which is absent in our phase diagram. This shows that even for broad Feshbach resonances it is essential to explicitly include the bosonic degree of freedom when working in an optical lattice.

%%%%%%%%%%%%%%%%%%%%%%%%%%%%%%%%%%%%%%%%%%%%%%%%%%%%%%%%%%%%%%%%%%%%%%%%%%%%%%%%%%%%%%%%%%%%%%%
\begin{figure}[hbpt]
\begin{center}
\subfigure[]{
\label{sc_BT_BCS}
\begin{minipage}[b]{0.45\textwidth}
\centering \includegraphics[width=1.0\textwidth]{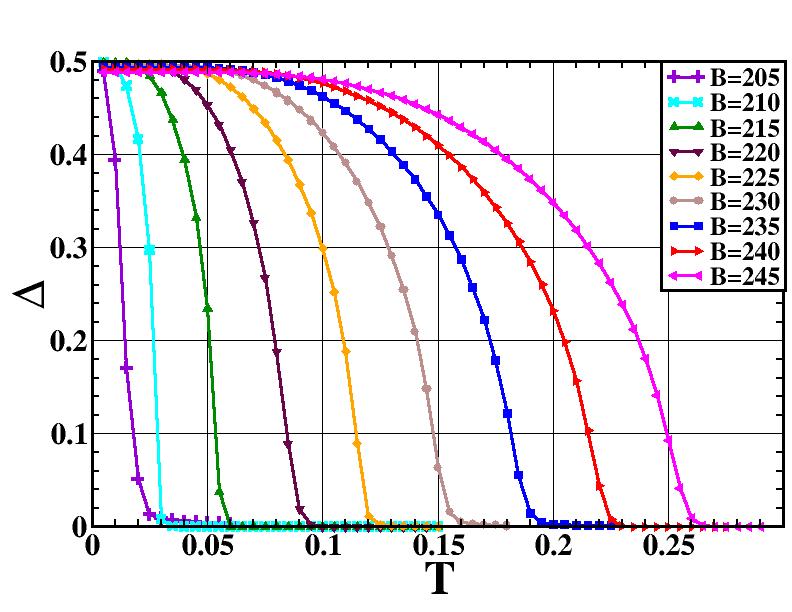}
\end{minipage}}
\hspace{0.25cm}
\subfigure[]{
\label{PD}
\begin{minipage}[b]{0.45\textwidth}
\centering \includegraphics[width=1.0\textwidth]{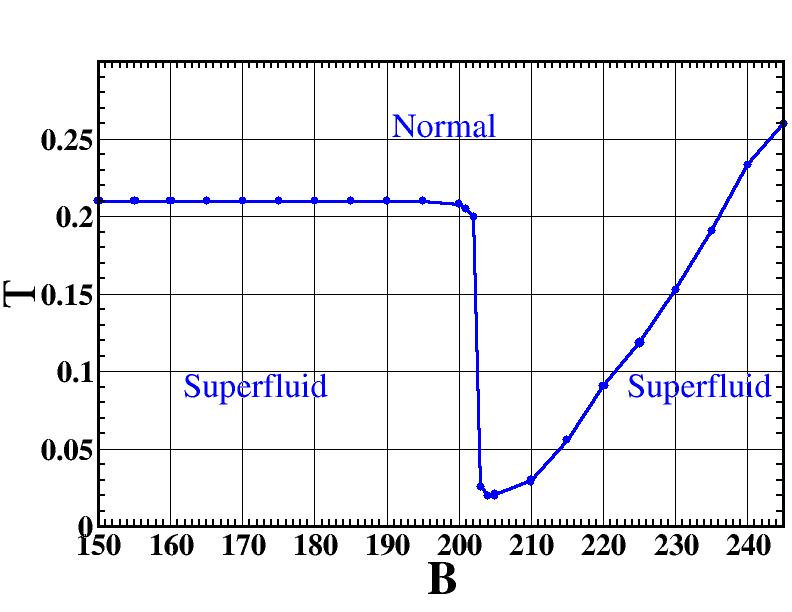}
\end{minipage}}\\
\caption{Finite temperature results: in subfigure (a)  we plot the fermionic superfluid order parameter as a function of temperature $T$ for different magnetic fields above the resonance, while in subfigure (b) we show the phase diagram. The blue solid line separates the superfluid phase from the normal phase. Here temperature is measured in units of the fermionic hopping $t_f$.}
%The number for fermions, bosons (sub-figure (a) ) and superfluid and superfluid order parameters (sub-figure (b) ) as a function of magnetic field $B$ for $T=0$. Full GDMFT calculation.}
\label{finite_T}
\end{center}
\end{figure}
%%%%%%%%%%%%%%%%%%%%%%%%%%%%%%%%%%%%%%%%%%%%%%%%%%%%%%%%%%%%%%%%%%%%%%%%%%%%%%%%%%%%%%%%%%%%%%%%%%%%%%%%
%%%%%%%%%%%%%%%%%%%%%%%%%%%%%%%%%%%%%%%%%%%%%%%%%%%%%%%%%%
%\begin{figure}
%\begin{center}
%\includegraphics[scale=0.35]{SC.jpg} 
%\end{center} 
%\vspace{-0.2cm}
%\caption{The superfluid order parameter as a function of temperature $T$ for different magnetic field above the resonance. Temperature is measured in units of the fermionic hopping $t_f$.}
%\label{sc_BT_BCS}
%\end{figure}
%%%%%%%%%%%%%%%%%%%%%%%%%%%%%%%%%%%%%%%%%%%%%%%%%%%%%%%%%%
%%%%%%%%%%%%%%%%%%%%%%%%%%%%%%%%%%%%%%%%%%%%%%%%%%%%%%%%%%
%\begin{figure}
%\begin{center}
%\includegraphics[scale=0.35]{PD.jpg} 
%\end{center} 
%\vspace{-0.2cm}
%\caption{The phase diagram. Blue solid line separate BEC/BCS phase from normal phase.}
%\label{sc_BT_BCS}
%\end{figure}
%%%%%%%%%%%%%%%%%%%%%%%%%%%%%%%%%%%%%%%%%%%%%%%%%%%%%%%%%%

\subsection{Nonzero temperature}
Having clarified the ground state phase diagram, we now consider finite temperature. In particular we investigate the critical temperature for the transition to the normal state. Deep in the BEC regime, the critical temperature is constant ($T_c \approx  0.21t_f$) and completely determined by the properties of the bosons: the bosonic hopping parameter $t_b$, the interbosonic background scattering length $a_b$ and the bosonic density. Only very close to resonance the critical temperature suddenly drops (see Fig. \ref{finite_T}). This coincides with the magnetic field value for which fermions enter the system. On the BCS side of the resonance, the critical temperature depends on the magnetic field and increases with $B$ (see Fig. \ref{finite_T}). This implies that at resonance the critical temperature is \emph{minimal}. This is in sharp contrast to the situation where no lattice is present, in which case the critical temperature is \emph{maximal} close to resonance.

This surprising fact can be understood from the behavior of the critical temperature in the single-band attractive Hubbard model \cite{Keller01, Toschi05}. In this model, the critical temperature is low both for very large and very small attraction and has a maximum in between. The reason for the low critical temperature at small attraction is the conventional exponential suppression of $T_c$ in the BCS regime. For strong attraction the critical temperature decreases again because the fermions start forming bound pairs with a greatly enhanced effective mass. Identifying the resonance position with the case of very large attraction, this explains the low critical temperature at this point. When moving away from resonance, the effective attraction induced by the Feshbach resonance becomes weaker and hence the critical temperature increases again. Far away from resonance one would therefore expect to find a maximum of the critical temperature, after which it decreases again because the BCS regime of weak attraction is entered.  However, due to the transition into the Mott insulator phase for the unit filling considered here  we cannot see the maximum of the critical temperature. Estimating the induced attractive interaction at the transition point to the Mott insulator by assuming it to be equal to the repulsive background interaction, this indeed gives a value for the induced attractive interaction which is larger (in absolute value) than the position of the maximum in the single-band attractive Hubbard model  \cite{Keller01, Toschi05}.

Our calculation also shows that on both sides of the resonance the ratio $\langle c_{\uparrow} c_{\downarrow}\rangle/\langle b^\dagger \rangle$ as a function of the temperature for fixed value of the magnetic field is constant. This means that the on-site Bose-Fermi coherence is not affected by the temperature for the low temperatures considered here.

%%%%%%%%%%%%%%%%%%%%%%%%%%%%%%%%%%%%%%%%%%%%%%%%%%%%%%%%%%
%\begin{figure}
%\includegraphics[scale=0.475]{b.png}  
%\vspace{-0.2cm}
%\caption{The superfluid order parameter as a function of magnetic field $B$ and temperature $T$. Single site calculation. \\
%{ comment: Temperature measured in units of $J$, later on I will change it.}}
%\label{b_BT}
%\end{figure}
%%%%%%%%%%%%%%%%%%%%%%%%%%%%%%%%%%%%%%%%%%%%%%%%%%%%%%%%%%

\section{Summary}\label{Summary}

We have studied ultracold fermionic $^{40}{\rm K}$ atoms in a three-dimensional optical lattice close to a Feshbach resonance. We derived an effective description in terms of a Bose-Fermi Hubbard model, in which the molecular degree of freedom is explicitly present. Our calculations show in agreement with Ref. \cite{Koetsier} that the effect of higher bands is crucial for a correct description of the Feshbach physics. We therefore take into account the fermionic occupation of higher bands.

To solve the strongly interacting multi-band problem we decouple the higher bands from the lowest one via a mean-field decoupling and reduce the Hamiltonian to an effective single-band Bose-Fermi Hubbard model which is self-consistently coupled to the higher bands. To solve this resulting model we use GDMFT.

The low temperature physics close to the Feshbach resonance is very rich. Upon changing of the magnetic field, the ratio of fermionic and bosonic densities is changing. Below the resonance the system is mainly occupied by molecular bosons forming a condensate. Close to  resonance the number of bosons decreases while the number of fermions increases. The fermions are in the superfluid phase. This resembles the BEC/BCS crossover close to a Feshbach resonance without an optical lattice. In addition, for the unit total filling considered here we found a transition into the fermionic Mott insulating phase when the magnetic field is increased even further. The Mott insulator phase is stabilized by the repulsive fermionic background scattering, which at large magnetic fields overcomes the attractive interaction induced by the Feshbach resonance. The phase transition into the Mott insulator is found to be of first order. We found that higher bands are crucial for a quantitatively correct prediction of the transition point.

We also calculated the critical temperature of the BEC/BCS superfluid phase across the resonance. Below resonance the critical temperature is independent of the magnetic field, until it sharply drops close to resonance. Above the resonance the critical temperature is increasing again, leading to the remarkable result of a \emph{minimal} critical temperature at resonance.

\section*{ACKNOWLEDGMENTS}
We thank H.T.C. Stoof and A. Koetsier for useful discussions.
This work was supported by the German Science Foundation DFG via grant \mbox{HO 2407/2-1}, the Collaborative Research Center SFB/TR 49 and the Nederlandse Organisatie for Wetenschappelijk Onderzoek NWO.  

\appendix
\section{Derivation of the self-energy for the Bose-Fermi mixture}\label{Appendix}

 In this appendix we evaluate the self-energy via correlation functions.  For this purpose we use the equation of motion, which in general has the following form:
%%%%%%%%%%%%%%%%%%%%%%%%%%%%%%%%%%%%
\begin{eqnarray}
\label{Eq_moution_A.SE}
i\omega_{n} \langle\langle \hat A, \hat B \rangle\rangle_\omega +
\langle\langle\left[\hat{\mathcal H}, \hat A \right]_{-},\hat B \rangle\rangle_\omega=
\langle \left[\hat A,\hat B\right]_{\eta}\rangle  \, .
\end{eqnarray}
%%%%%%%%%%%%%%%%%%%%%%%%%%%%%% 
Here $\omega_{n}$ are the Matsubara frequencies and $\langle\langle \hat A, \hat B \rangle\rangle_\omega$ is the general form of the Green's function, with the usual notation $[\hat A,\hat B]_\pm \equiv \hat A \hat B \pm  \hat B \hat A$; the plus sign applies when both operators are fermionic, otherwise the minus sign is used.

For the resonantly interacting Bose-Fermi mixture the generalized single impurity Anderson Hamiltonian has the following form:
%%%%%%%%%%%%%%%%%%%%%%%%%%%%%%%%%%%%
\begin{eqnarray}
\label{Hamiltonian_A.SE}
&&\hspace{-1.25cm}\hat{\mathcal H}^{\rm And}=-\sum_\sigma \mu_{f\sigma}\hat n^{f}_\sigma+U_f \hat n_{\uparrow}^f \hat n_{\downarrow}^f+U_{fb}\hat n^f\hat n^b
+g\left(\hat f^{\dagger}_{\downarrow}\hat f^{\dagger}_{\uparrow}\hat b+h.c.\right)+\hat {\cal H}_{b}^{\rm And} \nonumber\\
&&\hspace{-0.75cm}+\sum_{k\sigma} V_{k\sigma}\left(\hat f^{\dagger}_\sigma \hat c_{k\sigma}^{\phantom\dagger}+h.c\right) +\sum_{k\sigma}\varepsilon_{k\sigma}\hat c_{k\sigma}^{\dagger}\hat c_{k\sigma}^{\phantom\dagger}
+\sum_{k} W_{k}\left(\hat c_{k\uparrow}^{\dagger} \hat c_{k\downarrow}^{\dagger}+h.c\right)\, ,
\end{eqnarray}
%%%%%%%%%%%%%%%%%%%%%%%%%%%%%% 
where $\hat f_\sigma^\dagger$ and $\hat c_{k\sigma}^{\dagger}$ are the fermionic creation operators on the ``impurity site'' and in the band respectively. $\hat b^\dagger$ is the bosonic creation operator on the impurity site. $ \hat n^{f}=\hat n_{\uparrow}^f+\hat n_{\downarrow}^f= \sum_\sigma \hat f^{\dagger}_\sigma \hat f_\sigma^{\phantom\dagger}$, $\hat n^{b}=\hat b^{\dagger}\hat b$ and  $\hat {\cal H}_{b}^{\rm And}$ is the bosonic part of the Hamiltonian.

To calculate the self-energy we first evaluate the following commutator relations
%%%%%%%%%%%%%%%%%%%%%%%%%%%%%%%%%%%%
\begin{eqnarray}
\label{commutation1_A.SE}
&&\hspace{-1cm}\left[ \hat{\mathcal H}^{\rm And},\hat f_\sigma\right]_{-}=
\mu_{f\sigma}\hat f_\sigma-U_f \hat f_\sigma \hat f^\dagger_{\bar\sigma}
\hat f_{\bar\sigma}^{\phantom\dagger}-U_{fb}\hat f_\sigma^{\phantom\dagger} \hat b^{\dagger}\hat b+
\sigma g \hat f^\dagger_{\bar\sigma} \hat b -\sum_{k}V_{k\sigma}\hat c_{k\sigma} \, ,\\
\label{commutation2_A.SE}
&&\hspace{-1cm}\left[ \hat{\mathcal H}^{\rm And},\hat f_\sigma^\dagger \right]_{-} = -\mu_{f\sigma}\hat f_\sigma^\dagger+
U_f\hat f_\sigma^\dagger \hat f^\dagger_{\bar\sigma}\hat f_{\bar\sigma}^{\phantom\dagger}
+U_{fb}\hat f_\sigma^{\dagger} \hat b^{\dagger}\hat b-\sigma g \hat f^{\phantom\dagger}_{\bar\sigma} \hat b^\dagger +
\sum_{k}V_{k\sigma}\hat c_{k\sigma}^\dagger \, ,\\
\label{commutation3_A.SE}
&&\hspace{-1cm}\left[ \hat{\mathcal H}^{\rm And},\hat c_{k\sigma}\right]_{-}=-\varepsilon_{k\sigma}\hat c_{k\sigma}
-V_{k\sigma}\hat f_\sigma -\sigma W_{k} \hat c^{\dagger}_{k\bar\sigma} \, ,\\
\label{commutation4_A.SE}
&&\hspace{-1cm}\left[ \hat{\mathcal H}^{\rm And},\hat c_{k\sigma}^{\dagger}\right]_{-}=\varepsilon_{k\sigma}\hat c_{k\sigma}^\dagger+V_{k\sigma}\hat f_\sigma^\dagger +\sigma W_{k} \hat c^{\phantom\dagger}_{k\bar\sigma} \, ,
\end{eqnarray}
%%%%%%%%%%%%%%%%%%%%%%%%%%%%%% 
where $\bar \sigma =-\sigma$.

Now we use the equation of motion (\ref{Eq_moution_A.SE}) for the case when  $\hat A=\hat f_\sigma$ and $\hat B=\hat f^{\dagger}_\sigma$. In combination with the commutation relation (\ref{commutation1_A.SE}) we  get:
%%%%%%%%%%%%%%%%%%%%%%%%%%%%%%%%%%%%
\begin{eqnarray}
\label{Eq1_A.SE}
&&\hspace{-1cm}\left(i\omega_{n}+\mu_{f\sigma}\right)\langle\langle \hat f_\sigma,\hat f^{\dagger}_\sigma \rangle\rangle_\omega-
U_{f}\langle\langle \hat f_\sigma^{\phantom\dagger} \hat f^{\dagger}_{\bar \sigma}\hat f_{\bar\sigma}^{\phantom\dagger},
\hat f^{\dagger}_\sigma\rangle\rangle_\omega \nonumber \\
&&-U_{fb}\langle\langle \hat f_\sigma^{\phantom\dagger} \hat b^{\dagger}\hat b,\hat f^{\dagger}_\sigma\rangle\rangle_\omega+\sigma g \langle\langle \hat f^\dagger_{\bar\sigma} \hat b,\hat f_\sigma^\dagger \rangle\rangle_\omega 
-\sum_{k}V_{k\sigma}\langle\langle \hat c_{k\sigma}^{\phantom\dagger},\hat f^{\dagger}_\sigma\rangle\rangle_\omega =1\, . 
\end{eqnarray}
%%%%%%%%%%%%%%%%%%%%%%%%%%%%%% 
To calculate $\langle\langle \hat c_{k \sigma},\hat f^{\dagger}_\sigma \rangle\rangle_\omega$ we again use the equation of motion 
Eq. (\ref{Eq_moution_A.SE}), but in this case with  $\hat A=\hat c_{k\sigma}$ and $\hat B=\hat f^{\dagger}_\sigma$. With Eq. (\ref{commutation3_A.SE}) we obtain the following relation:
%%%%%%%%%%%%%%%%%%%%%%%%%%%%%%%%%%%%
\begin{eqnarray}
\label{Eq2_A.SE}
\left(i\omega_{n}-\varepsilon_{k\sigma}\right)\langle\langle \hat c_{k\sigma}^{\phantom\dagger},
\hat f^{\dagger}_\sigma\rangle\rangle_\omega- 
V_{k\sigma }\langle\langle \hat f_\sigma^{\phantom\dagger},\hat f^{\dagger}_\sigma\rangle\rangle_\omega 
-\sigma W_{k} \langle\langle \hat c^{\dagger}_{k\bar\sigma},\hat f^{\dagger}_\sigma\rangle\rangle_\omega =0 \, .
\end{eqnarray}
%%%%%%%%%%%%%%%%%%%%%%%%%%%%%% 
Finally to calculate $\langle\langle \hat c_{k\bar\sigma}^{\dagger},\hat f^{\dagger}_\sigma\rangle\rangle_\omega$ we use Eq. (\ref{Eq_moution_A.SE}) with $\hat A=\hat c_{k\bar\sigma}^\dagger$ and $\hat B=\hat f^{\dagger}_\sigma$, which  results in
%%%%%%%%%%%%%%%%%%%%%%%%%%%%%%%%%%%%
\begin{eqnarray}
\label{Eq3_A.SE}
\left(i\omega_{n}+\varepsilon_{k\bar \sigma}\right)\langle\langle \hat c_{k\bar\sigma}^{\dagger},
\hat f^{\dagger}_\sigma\rangle\rangle_\omega+
V_{k\bar\sigma}\langle\langle \hat f_{\bar\sigma}^{\dagger},\hat f^{\dagger}_\sigma\rangle\rangle_\omega 
-\sigma W_{k} \langle\langle \hat c^{\phantom\dagger}_{k\sigma},\hat f^{\dagger}_\sigma\rangle\rangle_\omega =0\, .
\end{eqnarray}
%%%%%%%%%%%%%%%%%%%%%%%%%%%%%% 
From Eqs. (\ref{Eq2_A.SE}) and (\ref{Eq3_A.SE})  we derive
%%%%%%%%%%%%%%%%%%%%%%%%%%%%%%%%%%%%
\begin{eqnarray}
\label{Eq4_A.SE}
\hspace{0cm}\langle\langle \hat c_{k\sigma}^{\phantom\dagger},\hat f^{\dagger}_\sigma\rangle\rangle_\omega=
\frac{V_{k\sigma}(i\omega_{n}+\varepsilon_{k\bar\sigma})}{(i\omega_{n}-\varepsilon_{k\sigma}) (i\omega_n+\varepsilon_{k\bar\sigma})-W_{k}^2}\langle\langle \hat f_\sigma^{\phantom\dagger},\hat f^{\dagger}_\sigma\rangle\rangle_\omega \nonumber\\
\hspace{1cm}-\frac{\sigma V_{k\bar\sigma}W_{k}}{(i\omega_{n}-\varepsilon_{k\sigma})(i\omega_{n}+\varepsilon_{k\bar\sigma}) -W_{k}^2}\langle\langle \hat f_{\bar\sigma}^{\dagger},\hat f^{\dagger}_\sigma\rangle\rangle_\omega \, . 
\end{eqnarray}
%%%%%%%%%%%%%%%%%%%%%%%%%%%%%% 
Now we  combine Eq. (\ref{Eq4_A.SE}) with Eq. (\ref{Eq1_A.SE}) and obtain:
%%%%%%%%%%%%%%%%%%%%%%%%%%%%%% 
\begin{eqnarray}
\label{Eq5_A.SE}
&&\hspace{0cm}\left(i\omega_{n}+\mu_{f\sigma}-\Delta_\sigma(i\omega_{n})\right)G_\sigma(i\omega_n)-\Delta_{SC}(\sigma i\omega_n)F^{*}(-\sigma i\omega_n)
\nonumber\\
&&\hspace{1cm}-U_f Q_{ff\sigma}(i\omega_n)-U_{fb} Q_{fb\sigma}(i\omega_n)-\sigma g Q_{g \bar\sigma\sigma}^{*}(i\omega_n)=1 \, .
\end{eqnarray}
%%%%%%%%%%%%%%%%%%%%%%%%%%%%%%
Note that $\langle\langle \hat f^{\phantom\dagger}_\sigma,\hat f^{\dagger}_\sigma \rangle\rangle_\omega\equiv G_\sigma(i\omega_n)$ is the normal Green's function  and $\langle\langle \hat f_{\uparrow}^{\phantom\dagger},\hat f^{\phantom\dagger}_\downarrow \rangle\rangle_\omega\equiv F(\omega)$ the superfluid Green's function. We also define
%%%%%%%%%%%%%%%%%%%%%%%%%%%%%% 
\begin{eqnarray}
\label{Hybrid_n_A.SE}
\Delta_\sigma(i\omega_n)&=&\Delta_\sigma^*(-i\omega_n) 
=-\sum_{k}V_{k\sigma}^2\frac{i\omega_n+
\varepsilon_{k\bar\sigma}}{(\varepsilon_{k\sigma}-i\omega_{n})(\varepsilon_{k\bar\sigma}+i\omega_{n})+W_k^2}  \, ,\\
\label{Hybrid_SC_A.SE}
\Delta_{SC}(i\omega_n)&=&\Delta_{SC}^*(-i\omega_n)
=\sum_{k} \frac{V_{k\uparrow}  V_{k\downarrow}W_{k}}{(\varepsilon_{k\uparrow}-i\omega_{n}) (\varepsilon_{k\downarrow}+i\omega_{n})+W_{k}^2}  \, , 
\end{eqnarray}
%%%%%%%%%%%%%%%%%%%%%%%%%%%%%%
as the normal and the superfluid hybridization functions respectively, and the following correlation functions:
$Q_{ff\sigma}(i\omega_n)=\langle\langle \hat f^{\phantom\dagger}_\sigma \hat f^{\dagger}_{\bar\sigma} \hat f^{\phantom\dagger}_{\bar\sigma} ,\hat f^{\dagger}_\sigma \rangle\rangle_\omega$, 
$Q_{ff\sigma\bar\sigma}(i\omega_n)=\langle\langle \hat f^{\phantom\dagger}_\sigma \hat f^{\dagger}_{\bar\sigma} \hat f^{\phantom\dagger}_{\bar\sigma} , \hat f^{\phantom\dagger}_{\bar\sigma} \rangle\rangle_\omega $, 
$Q_{fb\sigma}(i\omega_n)=\langle\langle \hat f^{\phantom\dagger}_\sigma \hat b^{\dagger}\hat b, \hat f^{\dagger}_\sigma \rangle\rangle_\omega $, $Q_{fb\sigma\bar\sigma}(i\omega_n)=\langle\langle \hat f^{\phantom\dagger}_\sigma \hat  b^{\dagger}\hat b, \hat f^{\phantom\dagger}_{\bar\sigma} \rangle\rangle_\omega $, 
$Q_{g\sigma}(i\omega_n)=\langle\langle \hat f^{\phantom\dagger}_\sigma \hat b^{\dagger}, \hat f^{\dagger}_\sigma \rangle\rangle_\omega $ and $Q_{g\sigma\bar\sigma}(i\omega_n)=\langle\langle \hat f^{\phantom\dagger}_\sigma \hat b^{\dagger}, \hat f^{\phantom\dagger}_{\bar\sigma} \rangle\rangle_\omega$.
%%%%%%%%%%%%%%%%%%%%%%%%%%%%%% 
%\begin{eqnarray}
%\label{Qff_A.SE}
%&&Q_{ff\sigma}(i\omega_n)=\langle\langle \hat f^{\phantom\dagger}_\sigma \hat f^{\dagger}_{\bar\sigma} \hat f^{\phantom\dagger}_{\bar\sigma} ,\hat f^{\dagger}_\sigma \rangle\rangle_\omega \, ,~ %\quad\quad\quad\quad
%Q_{ff\sigma\bar\sigma}(i\omega_n)=\langle\langle \hat f^{\phantom\dagger}_\sigma \hat f^{\dagger}_{\bar\sigma} \hat f^{\phantom\dagger}_{\bar\sigma} , 
%\hat f^{\phantom\dagger}_{\bar\sigma} \rangle\rangle_\omega \, ,\nonumber \\
%\label{Qfb_A.SE}
%&&Q_{ff\sigma}(i\omega_n)=\langle\langle \hat f^{\phantom\dagger}_\sigma \hat b^{\dagger}\hat b ,\hat f^{\dagger}_\sigma \rangle\rangle_\omega \, ,~ %\hspace{1.05cm}\quad\quad
%Q_{ff\sigma\bar\sigma}(i\omega_n)=\langle\langle \hat f^{\phantom\dagger}_\sigma \hat b^{\dagger}\hat b, \hat f^{\phantom\dagger}_{\bar\sigma} \rangle\rangle_\omega \, ,  \\
%\label{Qg_A.SE}
%&&Q_{ff\sigma}(i\omega_n)=\langle\langle \hat f^{\phantom\dagger}_\sigma \hat b^{\dagger} ,\hat f^{\dagger}_\sigma \rangle\rangle_\omega  \, ,~ %\hspace{1.25cm}\quad\quad
%Q_{ff\sigma\bar\sigma}(i\omega_n)=\langle\langle \hat f^{\phantom\dagger}_\sigma \hat b^{\dagger}, \hat f^{\phantom\dagger}_{\bar\sigma} \rangle\rangle_\omega \, . \nonumber
%\end{eqnarray}
%%%%%%%%%%%%%%%%%%%%%%%%%%%%%%

To obtain the self-energy we need to derive one more equation. For this purpose, we again use the equation of motion Eq. (\ref{Eq_moution_A.SE}) and take $\hat A=\hat f_\sigma^\dagger$ and $\hat B=\hat f_{\bar\sigma}^\dagger$. Based on Eq. (\ref{commutation2_A.SE}) we get:
%%%%%%%%%%%%%%%%%%%%%%%%%%%%%%%%%%%%
\begin{eqnarray}
\label{Eq7_A.SE}
&&\hspace{-2cm}\left(i\omega_{n}-\mu_{f\sigma}\right)\langle\langle \hat f_\sigma^\dagger,\hat f_{\bar\sigma}^{\dagger} \rangle\rangle_\omega+
U_{f}\langle\langle \hat f_{\sigma}^{\dagger} \hat f^{\dagger}_{\bar\sigma}\hat f_{\bar\sigma}^{\phantom\dagger}, 
\hat f^{\dagger}_{\bar\sigma} \rangle\rangle_\omega 
+ U_{fb}\langle\langle \hat f_\sigma^{\dagger} \hat b^{\dagger}\hat b,
\hat f^{\dagger}_{\bar\sigma} \rangle\rangle_\omega-\sigma g \langle\langle \hat f^{\phantom\dagger}_{\bar\sigma} \hat b^\dagger,\hat f_{\bar\sigma}^\dagger \rangle\rangle_\omega \nonumber\\
&&\hspace{-0.5cm}+\sum_{k}V_{k\sigma}\langle\langle \hat c_{k\sigma}^{\dagger},
\hat f^{\dagger}_{\bar\sigma}\rangle\rangle_\omega =0\, .
\end{eqnarray}
%%%%%%%%%%%%%%%%%%%%%%%%%%%%%% 
We now replace $\langle\langle \hat c_{k\sigma}^{\dagger},\hat f^{\dagger}_{\bar\sigma}\rangle\rangle_\omega$ using Eq. (\ref{Eq4_A.SE}) and obtain:
%%%%%%%%%%%%%%%%%%%%%%%%%%%%%% 
\begin{eqnarray}
\label{Eq10_A.SE}
&&\hspace{-1cm}-\sigma\left(i\omega_n-\mu_{f\sigma}+\Delta_\sigma^*(i\omega_n)\right)F^{*}(\sigma i\omega_n) 
+\sigma\Delta_{SC}(-\sigma i\omega_n)G_{\bar\sigma}(i\omega_n)- U_f Q_{ff,\sigma\bar\sigma}^{*}(i\omega_n) \nonumber\\
&&-U_{fb} Q_{fb\sigma\bar\sigma}^{*}(i\omega_n)-\sigma g Q_{g\bar\sigma}(i\omega_n) =0 \, .
\end{eqnarray}
%%%%%%%%%%%%%%%%%%%%%%%%%%%%%%

We proceed to write our results in matrix form. For $\sigma=1$ we use Eq. (\ref{Eq5_A.SE}) and the complex conjugate of Eq. (\ref{Eq10_A.SE}), while for $\sigma=-1$ we take Eq. (\ref{Eq10_A.SE}) and the complex conjugate of Eq. (\ref{Eq5_A.SE}):
%%%%%%%%%%%%%%%%%%%%%%%%%%%%%% 
\begin{eqnarray}
\label{Eq11_A.SE}
&&\hspace{-1cm}\left(i\omega_{n}+\mu_{f\uparrow}-\Delta_\uparrow(i\omega_{n})\right)G_\uparrow(i\omega_n)-\Delta_{SC}(i\omega_n)F^{*}(-i\omega_n) \nonumber\\
&&-U_f Q_{ff\uparrow}(i\omega_n)-U_{fb} Q_{fb\uparrow}(i\omega_n)-g Q_{g\downarrow\uparrow}^{*}(i\omega_n)=1 \, , \nonumber\\
\label{Eq12_A.SE}
&&\hspace{-1cm}-\left(i\omega_{n}-\mu_{f\downarrow}+\Delta_\downarrow(-i\omega_{n})\right)G_\downarrow^*(i\omega_n)-\Delta_{SC}(i\omega_n)F(i\omega_n)
\nonumber\\
&&-U_f Q_{ff\downarrow}^*(i\omega_n)-U_{fb} Q_{fb\downarrow}^*(i\omega_n)+g Q_{g\uparrow\downarrow}(i\omega_n)=1 \, ,\nonumber\\
\label{Eq13_A.SE}
&&\hspace{-1cm}\left(i\omega_n+\mu_{f\uparrow}-\Delta_\uparrow(i\omega_n)\right)F(i\omega_n) +\Delta_{SC}(i\omega_n)G_{\downarrow}^*(i\omega_n) \nonumber\\
&&-U_f Q_{ff,\uparrow\downarrow}(i\omega_n)-U_{fb} Q_{fb\uparrow\downarrow}(i\omega_n)- g Q_{g\downarrow}^*(i\omega_n) =0 \, , \nonumber\\
\label{Eq14_A.SE}
&&\hspace{-1cm}\left(i\omega_n-\mu_{f\downarrow}+\Delta_\downarrow(i\omega_n)\right)F^{*}(i\omega_n)-\Delta_{SC}( i\omega_n)G_{\uparrow}(i\omega_n) \nonumber\\
&&-U_f Q_{ff,\downarrow\uparrow}^{*}(i\omega_n)-U_{fb} Q_{fb\downarrow\uparrow}^{*}(i\omega_n)+ g Q_{g\uparrow}(i\omega_n) =0 \, . \nonumber
\end{eqnarray}
%%%%%%%%%%%%%%%%%%%%%%%%%%%%%%

The last four equations can be rewritten in matrix form in the following way:
%\begin{widetext}
%%%%%%%%%%%%%%%%%%%%%%%%%%%%%% 
\begin{eqnarray}
\label{Eq15_A.SE}
&&\hspace{-3cm}\left(
\begin{array}{cc}
1&0\\
0 & 1
\end{array}
\right)=\left(
\begin{array}{cc}
i\omega_n+\mu_{f\uparrow}-\Delta_\uparrow(i\omega_n) & -\Delta_{SC}(i\omega_n) \\
-\Delta_{SC}(i\omega_n) & i\omega_n-\mu_{f\downarrow}+\Delta_\downarrow(-i\omega_n)
\end{array}
\right)\left(
\begin{array}{cc}
G_\uparrow(i\omega_n)& F(i\omega_n)\\
F^{*}(-i\omega_n) & -G_\downarrow^*(i\omega_n)
\end{array}
\right)  \\
&&\hspace{-3cm}-\left(
\begin{array}{cc}
U_f Q_{ff\uparrow}(i\omega_n)+U_{fb} Q_{fb\uparrow}(i\omega_n)+g Q_{g\downarrow\uparrow}^{*}(i\omega_n) &  
U_f Q_{ff,\uparrow\downarrow}(i\omega_n)+U_{fb} Q_{fb\uparrow\downarrow}(i\omega_n)+g Q_{g\downarrow}^*(i\omega_n)   \\
U_f Q_{ff,\downarrow\uparrow}^{*}(i\omega_n)+U_{fb} Q_{fb\downarrow\uparrow}^{*}(i\omega_n)-g Q_{g\uparrow}(i\omega_n)  &
U_f Q_{ff\downarrow}^*(i\omega_n)+U_{fb} Q_{fb\downarrow}^*(i\omega_n)-g Q_{g\uparrow\downarrow}(i\omega_n)
\end{array}
\right) \, . \nonumber  
\end{eqnarray}
%%%%%%%%%%%%%%%%%%%%%%%%%%%%%%
%\end{widetext}

Now we compare Eq. (\ref{Eq15_A.SE}) with the Dyson equation, which has the matrix form:
%%%%%%%%%%%%%%%%%%%%%%%%%%%%%% 
\begin{eqnarray}
\label{Dyson_A.SE}
\hat {\mathcal G}^{-1}(i\omega_n)-\hat \Sigma(i\omega_n)=\hat G^{-1}(i\omega_n) \, ,
\end{eqnarray}
%%%%%%%%%%%%%%%%%%%%%%%%%%%%%%
where 
%%%%%%%%%%%%%%%%%%%%%%%%%%%%%% 
\begin{eqnarray}
\label{GreenG_A.SE}
\hat G(i\omega) =\left(
\begin{array}{cc}
G_\uparrow(i\omega_n)& F(i\omega_n)\\
F^{*}(-i\omega_n) & -G_\downarrow^*(i\omega_n)
\end{array}
\right)  
\end{eqnarray}
%%%%%%%%%%%%%%%%%%%%%%%%%%%%%%
is the matrix interacting Green's function, 
%%%%%%%%%%%%%%%%%%%%%%%%%%%%%% 
\begin{eqnarray}
\label{WeissG_A.SE}
\hat {\mathcal G}(i\omega)=\left(
\begin{array}{cc}
i\omega_n+\mu_{f\uparrow}-\Delta_\uparrow(i\omega_n) & -\Delta_{SC}(i\omega_n) \\
-\Delta_{SC}(i\omega_n) & i\omega_n-\mu_{f\downarrow}+\Delta_\downarrow(-i\omega_n)
\end{array}
\right)^{-1}
\end{eqnarray}
%%%%%%%%%%%%%%%%%%%%%%%%%%%%%% 
is  the matrix Weiss Green's function and $\hat \Sigma(\omega)$ is the matrix self-energy. From this comparison it follows directly that

%\begin{widetext}
%%%%%%%%%%%%%%%%%%%%%%%%%%%%%% 
\begin{eqnarray}
\label{SE_Matrix_A.SE}
&&\hspace{-3cm}\left(
\begin{array}{cc}
\Sigma_\uparrow(i\omega_n)& \Sigma_{SC}(i\omega_n)\\
\Sigma_{SC}^*(i\omega_n) & -\Sigma_\downarrow^*(i\omega_n)
\end{array}
\right)\nonumber \\
&&\hspace{-3cm}=\left(
\begin{array}{cc}
U_f Q_{ff\uparrow}(i\omega_n)+U_{fb} Q_{fb\uparrow}(i\omega_n)+g Q_{g\downarrow\uparrow}^{*}(i\omega_n) &  
U_f Q_{ff,\uparrow\downarrow}(i\omega_n)+U_{fb} Q_{fb\uparrow\downarrow}(i\omega_n)+g Q_{g\downarrow}^*(i\omega_n)   \\
U_f Q_{ff,\downarrow\uparrow}^{*}(i\omega_n)+U_{fb} Q_{fb\downarrow\uparrow}^{*}(i\omega_n)-g Q_{g\uparrow}(i\omega_n)  &
U_f Q_{ff\downarrow}^*(i\omega_n)+U_{fb} Q_{fb\downarrow}^*(i\omega_n)-g Q_{g\uparrow\downarrow}(i\omega_n)
\end{array}
\right) \nonumber\\
&& \hspace{-3cm} \times \left(
\begin{array}{cc}
G_\uparrow(i\omega_n)& F(i\omega_n)\\
F^{*}(-i\omega_n) & -G_\downarrow^*(i\omega_n)
\end{array}
\right) ^{-1}  \, .
\end{eqnarray}
%%%%%%%%%%%%%%%%%%%%%%%%%%%%%%

From here we obtain the final result:
%%%%%%%%%%%%%%%%%%%%%%%%%%%%%% 
\begin{eqnarray}
\label{SE_sigma.SE}
&&\hspace{-1.5cm}\Sigma_\sigma(i\omega_n)=\frac{\left(U_f Q_{ff\sigma}(i\omega_n)+U_{fb} Q_{fb\sigma}(i\omega_n)+\sigma g Q_{g\bar\sigma\sigma}^{*}(i\omega_n)\right) G_{\bar\sigma}^*(i\omega_n)}{G_\sigma(i\omega_n)G_{\bar\sigma}^*(i\omega_n)+F(\sigma i\omega_n)F^{*}(\bar\sigma i\omega_n)}\nonumber\\
&&+\frac{\left(\sigma U_f Q_{ff,\sigma\bar\sigma}(i\omega_n)+\sigma U_{fb} Q_{fb\sigma\bar\sigma}(i\omega_n)+g Q_{g\bar\sigma}^*(i\omega_n) \right) F^{*}(\bar\sigma i\omega_n)}{ G_\sigma(i\omega_n)G_{\bar\sigma}^*(i\omega_n)+F(\sigma i\omega_n)F^{*}(\bar\sigma i\omega_n) } \, ,\\
\label{SE_SC.SE}
&&\hspace{-1.5cm}\Sigma_{SC}(i\omega_n)=\frac{\left(U_f Q_{ff\uparrow}(i\omega_n)+U_{fb} Q_{fb\uparrow}(i\omega_n)+g Q_{g\downarrow\uparrow}^{*}(i\omega_n)\right) 
F(i\omega_n)}{G_\uparrow(i\omega_n)G_\downarrow^*(i\omega_n)+F(i\omega_n)F^{*}(-i\omega_n) }\nonumber\\
&&-\frac{\left(U_f Q_{ff,\uparrow\downarrow}(i\omega_n)+U_{fb} Q_{fb\uparrow\downarrow}(i\omega_n)+g Q_{g\downarrow}^*(i\omega_n)\right) G_{\uparrow}(i\omega_n)}{G_\uparrow(i\omega_n)G_\downarrow^*(i\omega_n)+F(i\omega_n)F^{*}(-i\omega_n) } \, ,% \\ 
\label{SE_SC_star.SE}
%&&\Sigma_{SC}^*(i\omega_n)=\frac{\left(U_f Q_{ff\downarrow}^*(i\omega_n)+U_{fb} Q_{fb\downarrow}^*(i\omega_n)-g Q_{g\uparrow\downarrow}(i\omega_n) \right) F^*(-i\omega_n)}{G_\uparrow(i\omega_n)G_\downarrow^*(i\omega_n)+F(i\omega_n)F^{*}(-i\omega_n) } \\
%&&\hspace{1.5cm}+\frac{\left(U_f Q_{ff,\downarrow\uparrow}^{*}(i\omega_n)+U_{fb} Q_{fb\downarrow\uparrow}^{*}(i\omega_n)-g Q_{g\uparrow}(i\omega_n) \right) G_\downarrow^*(i\omega_n)}{G_\uparrow(i\omega_n)G_\downarrow^*(i\omega_n)+F(i\omega_n)F^{*}(-i\omega_n) }
\, . %\nonumber
\end{eqnarray}
%%%%%%%%%%%%%%%%%%%%%%%%%%%%%%
%\end{widetext}

\end{document}